\documentclass{article}
\usepackage{caption}
\usepackage{subcaption}
\usepackage[utf8]{inputenc}
\usepackage[T1]{fontenc}
\usepackage{amsthm}
\usepackage{amsmath,amssymb}
\usepackage{amsfonts}
\usepackage{graphicx}
\usepackage{subcaption}
\usepackage{wrapfig} 
\usepackage{url}
\usepackage{color, xcolor}
\usepackage{titlesec}
\usepackage{multicol}
\usepackage{cite}
\usepackage{xcolor}
\usepackage{tabularx}
\usepackage{siunitx}
\usepackage{amsmath}
\usepackage{array}
\usepackage{comment}
\usepackage{hyperref} 
\usepackage{jcappub}
\usepackage{tikz}
\def\vs{\nonumber\\}
\usepackage{siunitx}
\hypersetup{
  colorlinks   = true,    
  urlcolor     = blue,    
  linkcolor    = blue,    
  citecolor    = blue      
}

\def\invMpc{\,h\,{\rm Mpc}^{-1}}

\title{BAO scale inference from biased tracers using the EFT likelihood}
\date{\today}

\author[a,b]{Ivana Babić,}
\author[b]{Fabian Schmidt,}
\author[b]{Beatriz Tucci}

\affiliation[a]{Arnold Sommerfeld Center for Theoretical Physics, Theresienstrasse 37, 80333 M\"unchen, Germany}
\affiliation[b]{Max–Planck–Institut f\"ur Astrophysik, Karl–Schwarzschild–Stra\ss e 1, 85748 Garching, Germany}

\emailAdd{i.babic@physik.uni-muenchen.de}
\emailAdd{fabians@mpa-garching.mpg.de}
\emailAdd{tucci@mpa-garching.mpg.de}

\abstract{
The physical scale corresponding to baryon acoustic oscillations (BAO), the size of the sound horizon at recombination, is precisely determined by CMB experiments.
Measuring the apparent size of the BAO scale imprinted in the clustering of galaxies gives us a direct estimate of the angular-diameter distance and the Hubble parameter as a function of redshift.
The BAO feature is damped by non-linear structure formation, which reduces the precision with which we can infer the BAO scale from standard galaxy clustering analysis methods.
Many methods to undo this damping via the so-called BAO reconstruction have so far been proposed; however, they all rely on backward modeling. In this paper, we present the first results of isotropic BAO inference from rest-frame halo catalogs using forward modeling combined with the EFT likelihood, in the case where the initial phases of the density field are fixed. We show that the remaining systematic bias is less than 2\% when we consider cutoff values of $\Lambda \leq 0.25 \invMpc$ for all halo samples considered, and below 1\% and consistent with zero for all but the most highly biased samples. 
We also demonstrate that, when compared to the standard power spectrum likelihood approach under the same assumption of fixed phases, the 1$\sigma$ errors associated to the field level inference of the BAO scale are 1.1 to 3.3 times smaller, depending on the value of the cutoff and the halo sample. Our analysis therefore unveils another promising feature of using field-level inference for high-precision cosmology.}

\begin{document}

\maketitle

\section{Introduction}

Baryon acoustic oscillations (BAO) are an oscillatory feature in the matter power spectrum. The same feature is visible in the correlation function as a bump located at the characteristic BAO scale $r_{s}$. 
The origin of the BAO can be found in the early Universe era when photons and baryons were tightly coupled by Compton scattering, forming the baryon-photon fluid. During this era, the gravitational force acting on the baryon perturbations was balanced by the radiation pressure resulting in acoustic oscillations of the baryon-photon fluid \cite{Eisenstein_1998_transfer}.  As the Universe expands and cools down, photons decouple. Traces of these sound waves remain visible as the acoustic oscillations in the CMB temperature anisotropies with the characteristic scale of the sound horizon at recombination, $r_s$. Essentially the same scale is imprinted in the acoustic density perturbations in the baryon distribution. Since baryons are coupled to dark matter gravitationally, and both jointly evolve under gravitational evolution after decoupling, the imprint of these early-time oscillations is visible at fixed comoving scale in the late-time clustering of matter. 
Given that the size of the sound horizon at  recombination has been well measured through CMB experiments, determining its apparent size in the late-time matter distribution allows us to estimate the angular-diameter distance and the Hubble parameter as a function of redshift. For a more detailed review of the BAO method see \cite{Weinberg_2013_review, Eisenstein_2005_review}.

Before we can apply this method however, we have to face the problem that 1) matter evolved nonlinearly; 2) we do not directly observe the evolved matter density field, but rather biased tracers of this field such as galaxies, galaxy clusters, quasars and others. 
The distribution of such objects at low redshifts is affected by the highly non-linear structure formation, both of the matter distribution itself and of the formation of the tracers themselves (see \cite{bernardeau/etal:2001} and \cite{Desjacques_2018_bias} for reviews on these topics, respectively). 
For the BAO feature specificially, nonlinear structure formation shifts and broadens the peak in the correlation function, and equivalently dampens the oscillations in the power spectrum on small scales \cite{Eisenstein_2007_modeling,padmanabhan/etal:2009,sherwin/zaldarriaga}. These effects reduce the precision with which the BAO can be measured from galaxy clustering \cite{Seo_2007_Fisher} by relying only on information available in the power spectrum.
As it was shown in \cite{Eisenstein_2007_modeling}, the dominant source of the broadening comes from bulk flows which are induced by large-scale modes, meaning that these effects can potentially be reversed. Therefore, there has been much interest in the BAO reconstruction methods \cite{Eisenstein_2007_reconstruction, Noh_2009_reconstruction,  Tassev_2012_reconstruction, Burden_2015_reconstruction, Schmittfull_2015_reconstruction, Wang_2017_reconstruction, Schmittfull_2017_reconstruction}. These methods start by smoothing the galaxy density field to filter out high-$k$ non-linearity, then this density is used to estimate the displacement field. Finally, the estimated displacement field is used to take tracers back to their estimated initial positions. Thus,  all these methods rely on a reverse or backward modeling approach. In addition, they have to make assumptions about galaxy bias and the cosmological model to infer the displacement field, rather than inferring all parameters jointly with the BAO scale.

In this paper we instead use the forward model approach to test
how well we can constrain the BAO scale by starting from the initial conditions. Forward modeling has gained a lot of momentum in the past few years \cite{1994Wiener, 1995, 1999Density, 20042dF, 2010_BayesianPS, 2010_FastHSampling, 2010_nonlinInference, 2010Recovering, 2011Multiscale,2013_icon_lss, 2014EUCLID, 2017SDSS, 2017Seljak, 2019Schmittfull, 2019Modi, ramanah/etal:2019}. 
One of the main advantages of this approach is in the fact that it does not rely on the correlation functions, instead it exploits the amplitudes and the phases of the tracer field directly. This is done by writing down a joint posterior for the initial density field, cosmological parameters and nuisance parameters (bias parameters and stochastic amplitudes). A crucial ingredient in this posterior is the likelihood function of observing a tracer field $\delta_{h}$ given the evolved matter density field. A likelihood function in the context of the effective field theory (EFT) of large scale structures \cite{Baumann_EFT, Carrasco_EFT} has been derived in \cite{Schmidt_2019_eft, sigma8_cosmology, Cabassa_2020}. 
A natural part of every EFT theory is a cutoff scale $\Lambda$ which corresponds to the maximum wavenumber of modes included in the calculations. The precise value of $\Lambda$ is arbitrary and its role is similar to the one of $k_{\rm{max}}$ in standard power spectrum analyses. Crucially, the results of all measurements should be independent of $\Lambda$. A natural upper limit for $\Lambda$ in the case of the EFT of LSS is the nonlinearity scale ($\Lambda_{\text{NL}} \approx 0.25 \invMpc$ at $z=0$), where perturbation theory of LSS breaks down. If this condition is satisfied, a controlled inference of the cosmological parameters and initial conditions can be performed. The result of applying this EFT likelihood to $\sigma_{8}$ inference has been presented in \cite{Schmidt_sigma8, sigma-eight-real}.  
A significant feature of the EFT likelihood is that it allows us to constrain the parameter of interest at the field level. Furthermore, 
Ref. \cite{Schmidt_2019_eft} has shown that the forward model combined with the EFT likelihood naturally includes the BAO reconstruction. 
In this paper, we follow up and perform an unbiased inference of the isotropic BAO scale, which we will refer to simply as BAO scale, from rest-frame halo catalogs using the EFT likelihood. Our tests are based on simulations, which enables us to fix the initial phases of the linear density field to their correct values (to avoid any possible misunderstanding, fixing the phases of the initial density field means both fixing its amplitude and phase in each grid voxel). This removes cosmic variance as much as possible and reduces the size of the error bars.

This paper is organized in the following way. In Sec. \ref{About eft} we give a short summary of the most important properties of the EFT likelihood. In Sec. \ref{sec: method} we give an overview of the method used for the inference of the BAO scale in the EFT based approach. 
In Sec. \ref{sec: eft results}, we use the field-level EFT likelihood to find the BAO scale value.
To gauge what are the improvements from the field level likelihood over standard, power-spectrum-based approaches, we also determine the BAO scale value using a likelihood constructed from the power spectrum (Sec. \ref{sec: PS results}). For the predicted power spectrum we do not perform any additional BAO reconstruction; instead, we use the deterministic halo field found from the forward model. 
This means that the EFT likelihood is still used to constrain the bias parameters in the construction of the predicted power spectrum, but is not used to constrain the BAO scale. On the other hand, our power spectrum covariance takes into account that the phases are fixed to the ground truth, i.e. there is no cosmic variance. Thus, we perform a fair comparison between both methods.
We conclude in Sec. \ref{sec: Conclusion}.

\section{The EFT Likelihood}\label{About eft}

We begin with a brief review of the EFT likelihood.
Throughout this paper we refer to the tracers considered as halos, simply because we are working with halo catalogs from simulations. However, since the EFT approach only assumes that the formation of the tracer is spatially local, all the results are equally applicable to galaxies or any other cosmological tracer. 
With $\delta_{h}$ we will denote the observed fractional number density perturbation  of a given halo sample. In the rest frame of a halo, this density field is given by
\begin{equation}
    \delta_{h}(\textbf{x}, \tau)  \equiv
    \frac{ n_{h}(\textbf{x},\tau) - \Bar{n}_{h}(\tau)}{\Bar{n}_{h}(\tau)},
\end{equation}
where $\tau$ is the conformal time,  $n_{h}(\textbf{x}, \tau)$ denotes the comoving rest-frame halo density and $\Bar{n}_{h}(\tau)$ is its position-independent mean.
In \cite{Schmidt_2019_eft} a joint posterior for the initial density field $\delta_{\rm in}$, cosmological parameters $\theta$ and nuisance parameters (bias parameters $b_{O}$ and stochastic amplitudes $\sigma_a$), $P(\delta_{\rm in}, \theta, b_{O}, \sigma_{a}|\delta_{h})$, was introduced. All the important physics of halo formation is contained within the likelihood $P(\delta_{h}|\delta_{\rm in}, \theta, {b_{O}, \sigma_{a}})$ giving the probability of observing the halo density $\delta_{h}$ given the initial conditions, cosmological and nuisance parameters. Once the initial density field $\delta_{\rm in}$ has been specified, there are three important parts of $P(\delta_{h}|\delta_{\rm in}, \theta, {b_{O}, \sigma_{a}})$ we need to focus on. Those are the deterministic forward model for matter, the bias relation and the  conditional likelihood for finding a measured halo density field given the matter density field and bias parameters. In the following we summarize the most important information about each of them. 

The deterministic forward model for matter $\delta = \delta_{\rm fwd}[\delta_{\rm in}]$ used in this paper is third-order Lagrangian perturbation theory (3LPT), as explained below. All initial perturbations with wavenumber $ k > \Lambda$, where $\Lambda$ is the initial cut-off, are set to zero. Further, we use the Lagrangian bias expansion
\begin{equation}\label{eq: bias expansion}
        \delta^{L}_{h,{\rm det}}(\textbf{q}, \tau) = \sum_{O}b^{L}_{O}O^{L}(\textbf{q}, \tau),
\end{equation}
where $b^{L}_{O}$ and $O^{L}$ are the Lagrangian bias coefficients and operators, respectively, and $\textbf{q}$ is the Lagrangian coordinate marking the initial position of the particle as $\tau \xrightarrow[]{} 0$. 
In the Lagrangian bias expansion, we first construct the bias operators and then displace them to Eulerian frame. This can be done conveniently in the same step as the LPT calculations. The relationship between the Lagrangian position \textbf{q} and and the final Eulerian position of the matter particles $\textbf{x}$ is given through the displacement vector $\textbf{s}$,
\begin{equation}\label{eq: Lagr to euler}
    \textbf{x}(\tau) = \textbf{q} + \textbf{s}(\textbf{q}, \tau).
\end{equation}
In LPT, we treat the components of the displacement tensor as small parameters, which allows us to write
\begin{equation}
    \textbf{s}(\textbf{q}, \tau) = \sum_{n=1}^{\infty}s^{(n)}(\textbf{q}, \tau).
\end{equation}
This expansion of the displacement tensor is related to the expansion in powers of $\delta$ performed in Eulerian Perturbation Theory (EPT). This can be seen from the following  consideration. From Eq. \eqref{eq: Lagr to euler}, we find the Jacobian $J_{ij}$ to be 
\begin{equation}\label{eq: Jacobian}
    J_{ij} = \frac{\partial x_i}{\partial q_j}= \delta_{ij} + M_{ij}(\textbf{q}, \tau) ,
\end{equation}
where we have introduced the Lagrangian deformation tensor 
\begin{equation}
    M_{ij} = \partial_{q,i}s_{j}(\textbf{q}, t).
\end{equation} Then using the continuity relation together with  Eq. \eqref{eq: Jacobian}, we find the relationship between the deformation tensor and the density field
\begin{equation}\label{eq: deltaM}
    1 + \delta(\textbf{x}(\textbf{q}, \tau), \tau) = |\textbf{1} + \textbf{M}(\textbf{q}, \tau)|^{-1}.
\end{equation} 
From Eq. \eqref{eq: deltaM}, it is clear that at first order in perturbations
\begin{equation} \label{eq: recursion}
   M^{(1)}_{ij}(\textbf{q}) = \frac{\partial_{q,i} \partial_{q,j}}{\nabla^2_q} \delta^{(1)}(\textbf{q}).
\end{equation}

The basis of the Lagrangian set of operators can be conveniently expressed in terms of the symmetric part of the Lagrangian deformation tensor \cite{biased_tracers_time_evolution_2015}
\begin{equation}
    M_{ij}^{(n)} = \partial_{q,(i} s_{j)}^{(n)}(\textbf{q}, t).
\end{equation}
The antisymmetric part, $\partial_{q,[i}s_{j]}$, appears from third order in perturbations, but does not need to be included in the bias expansion as it is redundant \cite{biased_tracers_time_evolution_2015}. 
We work by treating the components of the deformation tensor as small parameters and find the Lagrangian operators at each perturbative order by taking all the scalar contractions of $M_{ij}^{(n)}$. We do not need to include $\text{tr}[M^{(n)}]$ for $n>1$ since those can always be expressed in terms of scalars constructed using the lower order operators. Eq. \eqref{eq: recursion} is the starting point in a recursion relation that can be used to construct the tensors $M^{(n)}$ at all orders \cite{rec_LPT}.
In our calculations, we will be using the operators up to third order in perturbations which are listed here according to their perturbative order \cite{large_scale_bias_2018}
\begin{equation}\label{list bias}
    \begin{split}
        &1^{\text{st}}\quad \quad \delta, \nabla^{2}_x \delta \\
        &2^{\text{nd}}\quad \quad \text{tr}[(M^{(1)})^{2}], \text{ tr}[M^{(1)}]^{2}\\
        &3^{\text{rd}}\quad \quad \text{tr}[(M^{(1)})^{3}], \text{ tr}[M^{(1)}]^{3}, \text{ tr}[M^{(1)}M^{(2)}], \text{ tr}[M^{(1)}]^2 \text{tr}[M^{(1)}].
    \end{split}
\end{equation}\label{eq}
The corresponding bias coefficients are 
\begin{equation}
    \begin{split}
        &1^{\text{st}}\quad \quad b_{\delta}, b_{\nabla^{2} \delta}  \\
        &2^{\text{nd}}\quad \quad b_{\text{tr}[(M^{(1)})^{2}]}, b_{\text{ tr}[M^{(1)}]^{2}}\\
        &3^{\text{rd}}\quad \quad b_{\text{tr}[(M^{(1)})^{3}]}, b_{\text{ tr}[M^{(1)}]^{3}}, b_{\text{ tr}[M^{(1)}M^{(2)}]}, b_{\text{tr}[M^{(1)}]^2 \text{tr}[M^{(1)}]}.
    \end{split}
\end{equation}

With these ingredients, we are finally able to construct the Lagrangian bias expansion presented in Eq. \eqref{eq: bias expansion}. The set of Eulerian operators in turn is then obtained by displacing each of the Lagrangian operators via Eq. \eqref{eq: Lagr to euler}. We define a grid of $512^3$ cells in which $O^{L}(\textbf{q}_i)$ is set as the weight (or mass) of a ``particle'' at position $\textbf{q}_i$. With the aim of preventing noise generation on large scales, we then deposit the particle mass at its Eulerian position $\textbf{x}_i$ using a could-in-cell scheme, such that the total mass is guaranteed to be conserved. This displacement technique is performed for all Lagrangian bias operators (up to the desired order) with the exception of $\text{tr}[M^{(1)}]$. For the latter field, we instead displace a unity-weight field to obtain the Eulerian density, which is associated to the well-known Eulerian bias parameter $b_\delta$ commonly called $b_1$. Thus, we will hereafter use $b_{1} \equiv b_{\delta}$ following standard convention. More details about the implementation of this procedure can be found in \cite{sigma-eight-real}.

Note that the operator $\nabla^{2}_x \delta $ is not derived from the recursion relations arising from Eq. \eqref{eq: recursion}, but contains two more spatial derivatives. This operator is the leading \emph{higher-derivative} operator which accounts for the non-locality of halo formation. The coefficients of higher-derivative operators are thus related to the spatial scale $R_{*}$ which quantifies the size of the spatial region involved in the process of halo formation, and their contribution to $\delta_{h,{\rm det}}$ is suppressed by powers of $k^2 R_*^2$ on large scales.
  
Finally, we turn to the conditional likelihood which provides the probability for finding a measured halo density field given the matter density field and the bias parameters. The EFT likelihood which we use in this paper has the following form:
    \begin{equation}\label{eq: EFT likelihood}
        \ln P (\delta_{h}| \delta_{\rm in}, \{b_{O}\}) =
        -\frac{1}{2}\sum_{|k|<\Lambda}\Bigg[
        \ln [2\pi\sigma_{\varepsilon}^{2}(k)] + 
        \frac{1}{\sigma_{\varepsilon}^{2}(k)}
        |\delta_{h}(\textbf{k}) - \delta_{h, {\rm det}}[\delta_{\rm in}, \{b_{O}\}](\textbf{k}) |^{2}
        \Bigg].
    \end{equation}
    
The parametrization of  $\sigma^{2}$ is chosen in such a way to ensure that $\sigma^{2}$ is positive definite,
    \begin{equation}
       \sigma^2(k) = ( \sigma_{\varepsilon} + k^2\sigma_{\varepsilon,2} )^2.
    \end{equation}
We can interpret $\sigma_{\varepsilon}$ as the amplitude of halo stochasticity in the large-scale limit ($k \xrightarrow{} 0$). $\sigma_{\varepsilon, 2}$ is the leading scale-dependent correction to halo stochasticity and it captures the backreaction of small physical scales in real space. Since we found that $\sigma_{\varepsilon, 2}$ has a negligible contribution to our results, we set it to zero throughout the paper. 
A distinct feature of this likelihood is the existence of a hard cutoff $\Lambda$ which marks a boundary above which all the modes $k$ are integrated out. In other words, this cutoff ensures that we are focusing only on large scales. 
On such large scales, the central limit theorem guarantees that the noise fields for $k<\Lambda$ can be approximated to leading order as independent Gaussian degrees of freedom, making the resulting likelihood normal with diagonal covariance (see Fig.~8 of \cite{Schmidt_sigma8} for an explicit demonstration).
In our analysis we will be calculating the EFT likelihood for different cutoffs $\Lambda$ to see how the results of the BAO scale inference are influenced by different modes included in the calculation. Since the number of modes scales as $k^3$, and constraints are thus dominated by modes close to the cutoff $\Lambda$, the results for different $\Lambda$ values will be essentially independent.

For the results in Sec. \ref{sec: eft results}, we use the EFT likelihood to infer the value of the BAO scale. In this application, the likelihood is marginalized over all bias parameters as described in \cite{sigma8_cosmology}. The EFT likelihood is also used in Sec. \ref{sec: PS results} to find the values of the bias parameters $b_O$ to be used in the deterministic halo power spectrum. To find the value of a parameter $b_O$, we marginalize the EFT likelihood over all remaining bias parameters and then run the MINUIT minimizer \cite{minuit} to find the best fit for the $b_O$ parameter. We repeat this procedure for all the fields appearing in Eq. \eqref{list bias} (see also \cite{barreira/etal:2021,lazeyras/etal:2021} for a recent study of halo and galaxy bias using this method).

\section{Method}\label{sec: method}

The BAO is normally used to infer the angular diameter distance to a given observed redshift by comparing the predicted scale of the BAO feature to the data for a given assumed distance.
This kind of approach is not suitable in our case since we are working with simulations on a cubic box with periodic boundary conditions. To change the distances inside such a box, we would have to introduce a window function, and would not be able to keep the initial conditions fixed to the ground truth (since changing the fiducial distance amounts to changing the comoving volume of the data as well). In order to avoid these significant complications, we adopt a different approach, essentially rescaling the predicted comoving sound horizon.

\subsection{Approximating the power spectrum}\label{subsec: Approx_PS}

We want to constrain the BAO scale $r_s$ just from the information available in the oscillatory part of the power spectrum, without referring to its broad-band part. This is because the broad-band power spectrum depends on other cosmological parameters as well. 
One possible way to constrain $r_s$ from the power spectrum is by varying the baryon density $\omega_b$ and checking which value agrees the best with the data. However, varying the value of $\omega_b$ changes not only the oscillatory part of the power spectrum, but also its broad band. Therefore, this approach is not suitable. 

Instead, we approximate the linear matter power spectrum as
\begin{equation}\label{eq: power spectra full}
    P_{\rm L}(k, \beta) = P_{\rm L,sm}(k)[1+ A\sin (k\beta r_{\rm fid})\exp(-k/k_{\rm D}) ],
\end{equation}
where $A$ and $k_{\rm D}$ are constants and $r_{\rm fid}$
is the fiducial BAO scale. Through this equation we separate the broad band part of the power spectrum, described with the function $P_{\rm L,sm}(k)$, from its oscillatory feature. In the oscillatory feature we recognize the contribution $\sin(k\beta r_{\rm fid})$
describing the baryon acoustic oscillations and the exponential envelope corresponding to the primordial photon diffusion, or Silk damping. The Silk damping term absorbs all the physics that is not captured within the fluid approximation to the baryon-photon system before recombination.

Finally, we introduced the factor $\beta$ as
\begin{equation}
    \beta = \frac{r_{s}}{r_{\rm fid}}.
\end{equation}
By changing $\beta$, we are changing the size of the BAO scale $r_{s}$ to match the data while keeping the distances fixed. Most importantly, changing $\beta$ 
will result in changes in the oscillatory part of the power spectrum while keeping its overall shape intact. Notice that, since the BAO scale was imprinted in the power spectrum during the early Universe, varying it in the initial (linear) density field is the physically correct approach.

\begin{figure}
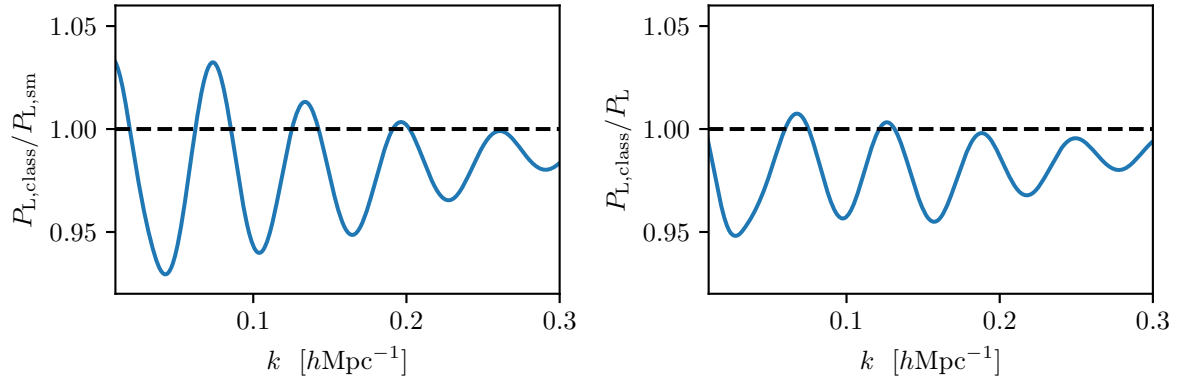

     \centering
     \begin{subfigure}[b]{0.49\textwidth}
         \centering
         \input{Figures_fit/Pklinear_over_Psmooth_PaperA.pgf}
        \caption{Ratio of the linear power spectrum obtained from CLASS and the best fit for $P_{\rm L,sm}$. }\label{Fig: psmooth}
     \end{subfigure}
     \hfill
     \begin{subfigure}[b]{0.49\textwidth}
         \centering
         \input{Figures_fit/Pk_bestFit_ratio_paperA_01.pgf} 
    \caption{Ratio of the linear power spectrum from CLASS and the best fit for $P_{\rm L}$ from Eq. \eqref{eq: power spectra full}.}\label{fig: Pk fit}
     \end{subfigure}
    \caption{Comparing the linear power spectrum to the power spectrum approximation. }
\end{figure}

The function $P_{L,sm}(k)$  can be written in the form
\begin{equation}\label{eq: Psmooth}
    P_{\rm L,sm}(k) = N \Big( \frac{k}{k_{\rm p}} \Big)^{n_{s}} T^{2}(k), 
\end{equation}
where $N$ is a normalisation constant that is proportional to the primordial normalization $\mathcal{A}_s$ times the growth factor squared, $k_{ \rm p}$ is the pivot scale and $T(k)$ is the ``no-wiggle'' transfer function which we take from Ref. \cite{Eisenstein_1998_transfer}. We found the value of $N$, $k_D$ and $A$ by fitting Eq. \eqref{eq: Psmooth} to the linear power spectrum produced by the CLASS code \cite{CLASS}. Fig. \ref{Fig: psmooth} shows the ratio of the CLASS power spectrum to $P_{\rm L,sm}(k)$. We can clearly see the damped oscillation in the BAO range which indicates that $P_{\rm L,sm}$ really does describe the smooth power spectrum with no BAO wiggles. Fig. \ref{fig: Pk fit} shows the ratio of the CLASS power spectrum to the power spectrum approximated by Eq. \eqref{eq: power spectra full}. While we do see some residual wiggles in the plot, we also notice that they are suppressed at high $k$ where most of the constraints come from. Therefore, we can conclude that Eq. \eqref{eq: power spectra full} is a good approximation of the linear power spectrum and we can use it for the BAO scale inference.

Given the known fiducial power spectrum, i.e. the power spectrum from which the initial conditions of the N-body simulations were drawn, it is easy to find the power spectrum with a different BAO scale, using  Eq. \eqref{eq: power spectra full}. We introduce the factor $f(k, \beta)$ as 
\begin{equation}
    f^{2}(k, \beta) = \frac{P_{\rm L}(k, \beta)}{P_{\rm fid}(k)} = \frac{1+ A\sin (k\beta r_{\rm fid})\exp(-k/k_{\rm D}) }{1+ A\sin (k r_{\rm fid})\exp(-k/k_{\rm D}) }. 
\end{equation}
Notice that $f(k,1) = 1$. 
From $f(k, \beta)$ it is straightforward to find the relationship between the fiducial and rescaled linear density fields 
\begin{equation}\label{eq: denisty}
   \delta_{\beta}(k, \beta) =  f(k, \beta) \delta_{\rm fid}(k).
\end{equation}
To recap, $\delta_\beta$ is the linear matter density field with all fiducial phases but for which the BAO scale is of the size $r_s = \beta r_{\rm fid}$. Throughout the paper we will be using different $\delta_\beta$ as the initial fields for our forward model. 

\subsection{Profile likelihood}

All numerical results presented here were obtained for a spatially flat $\Lambda$CDM cosmology with parameters $\Omega_{m} = 0.3$, $\Omega_{\Lambda} = 0.7$, $h = 0.7$, $n_{s}=0.967$ and a box with size $L = 2000 h^{-1}\text{Mpc}$. We use four halo mass bins in the mass range  $10^{12.5} h^{-1}  M_{\odot}$--$10^{14.5} h^{-1}  M_{\odot}$. We present results on two simulation realizations, ``run 1'' and ``run 2'', which differ in their initial phases. In Tab. \ref{eq: denisty}, we present the number density of halos in run 1 at different redshifts. 
\begin{table}
\centering
\begin{tabular}{c c c} 
 \hline
 \hline
 $z$ & Mass range &  $\bar n_h$ \\
 & [$\log_{10}(M/h^{-1}M_{\odot})$] & [$({\rm Mpc}/h)^{-3}$]  \\ 
 \hline
 \hline
 0.0 & $[12.5-13.0]$ & \num{7.05688 e-04} \\
 0.5 & $[12.5-13.0]$ & \num{6.00105e-04} \\ 
 1.0 & $[12.5-13.0]$ & \num{ 4.74483e-04}\\
 \hline
 0.0 & $[13.0-13.5]$ & \num{3.50970e-04 }\\
 0.5 & $[13.0-13.5]$ &  \num{2.76635e-04 }\\
 1.0 & $[13.0-13.5]$ & \num{1.88450e-04}\\
  \hline
 0.0 & $[13.5-14.0]$ & \num{1.14982e-04} \\
 0.5 & $[13.5-14.0]$ & \num{7.42665e-05}\\
 1.0 & $[13.5-14.0]$ & \num{3.76761e-05} \\
 \hline
 0.0 & $[14.0-14.5]$ & \num{2.96594e-05 }\\
 0.5 & $[14.0-14.5]$ & \num{1.33126e-05}\\
 1.0 & $[14.0-14.5]$ & \num{3.94175 e-06}\\ 
 \hline
 \hline
\end{tabular}
\caption{Number density of halos in run 1 at different redshifts.}\label{table: denisty}
\end{table}

\clearpage
\newpage

As mentioned earlier, we do not sample the initial density field; it is instead fixed to the exact initial conditions used in the N-body simulations within which the halos are identified.
To get the initial density with different BAO scales, we apply Eq. \eqref{eq: denisty} for a set of values $\{ \beta^i \}$. The default set spans the range  $[0.8, 1.02]$; in all cases, we make sure that the maximum-a-posteriori (MAP) value of $\beta$ is safely within the range. Fixing the initial phases not only saves the computational time, but it also minimizes the cosmic variance as much as possible resulting in smaller error bars for the inferred value of $\beta$.

To find the MAP estimate for $\beta$, which we denote as $\hat\beta$, we use the profile likelihood \cite{ProfileLikelihod}. For a probability distribution $P(\beta, \sigma_{\varepsilon}|\delta_{h})$ (recall that the bias coefficients are analytically marginalized over) and parameter $\beta$, the profile likelihood is defined as
\begin{equation}
    P^{\rm prof}(\beta) =
    \displaystyle{\max_{ \sigma_{\varepsilon}}}[P(\beta, \sigma_{\varepsilon}|\delta_{h})],
\end{equation}
where the parameter $\sigma_{\varepsilon}$ has been profiled out.

\begin{figure}[h!]
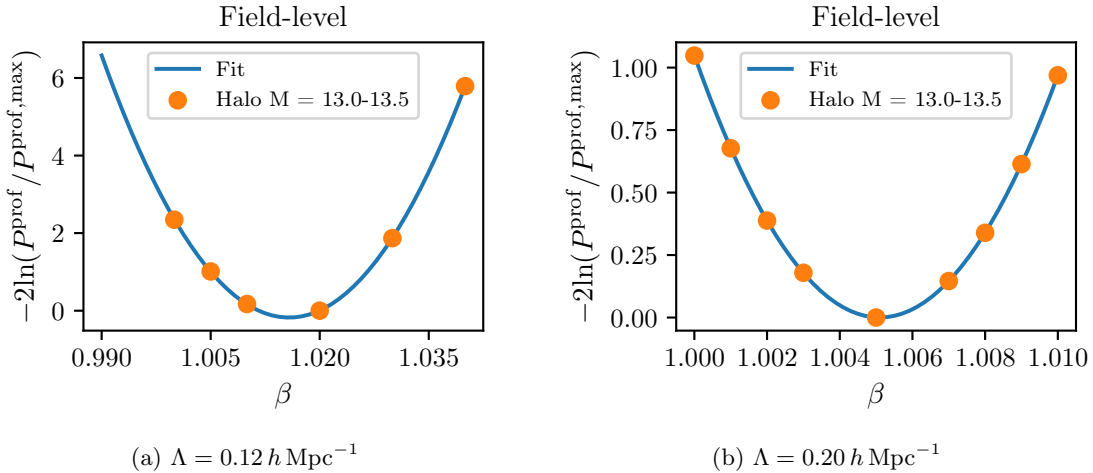

\centering
\begin{subfigure}{.5\textwidth}
  \centering
  \input{EFT_likelihood/Parabola_fit/Standard_Likelihood_beta_lambda=0.12.pgf}
  \caption{$\Lambda = 0.12 \invMpc$}
\end{subfigure}%
\begin{subfigure}{.5\textwidth}
  \centering
  \input{EFT_likelihood/Parabola_fit/Standard_Likelihood_beta_lambda=0.2.pgf}
  \caption{$\Lambda = 0.20 \invMpc$}
\end{subfigure}
\caption{Profile likelihood $-2\ln P^{\rm prof}$ plotted as a function of $\beta$ for two different cut-offs $\Lambda$ at $z=0$. The blue line shows the parabolic fit which was used to find the maximum-a-posteriori value $\hat{\beta}$ and its error $\sigma(\hat\beta)$.}\label{fig: parabola_eft}
\end{figure}

For a fixed $\Lambda$, halo sample, redshift and $\beta^{i}$ we maximize the profile likelihood using the MINUIT minimizer \cite{minuit}. In this way we obtain a set $\{\beta^{i}, -2\ln P^{\rm prof}(\beta^{i})  \}$ which is nicely fit by a parabola for all halo samples and all cutoffs. An example of this parabola for two different cutoffs is shown in  Fig. \ref{fig: parabola_eft}, where the elements of the set $\{\beta^{i}, -2\ln P^{\rm prof}(\beta^{i})  \}$ are represented with orange dots and the blue line corresponds to the parabolic fit.
The MAP value $\hat{\beta}$ is located at the minimum of the best fit parabola. The estimated 68\% confidence-level error on $\hat{\beta}$ is given by the inverse square root curvature of the parabolic fit.

\section{Field-Level Results} \label{sec: eft results}

\begin{figure}[h]
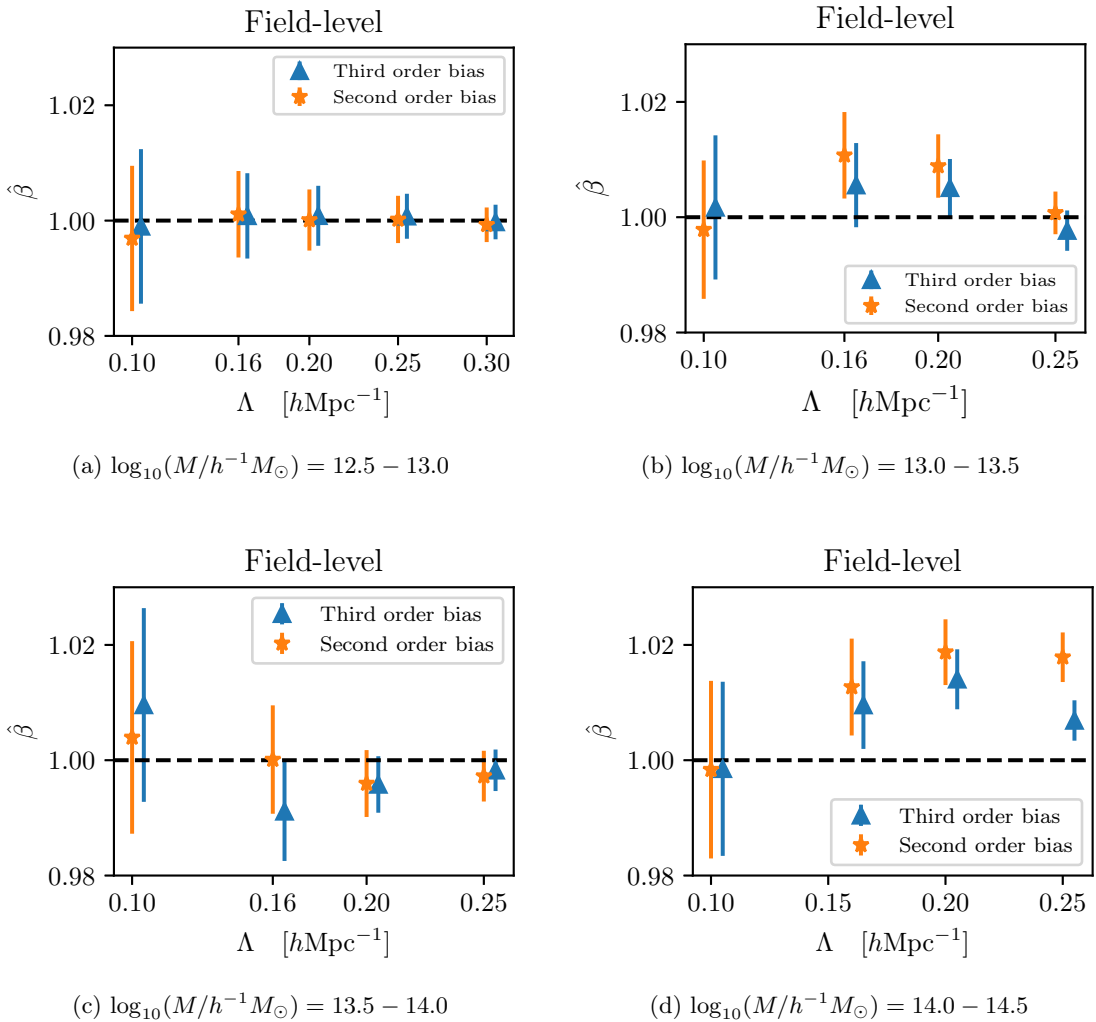

  \begin{subfigure}[b]{0.49\textwidth}
    \centering
    \input{EFT_likelihood/12.5-13/12.5-13.0_different_bias.pgf} 
    \caption{$\log_{10}(M/h^{-1}M_{\odot}) = 12.5-13.0$} 
    \vspace{4ex}
  \end{subfigure}
  \begin{subfigure}[b]{0.49\textwidth}
    \centering
    \input{EFT_likelihood/13_13.5/13.0-13.5_different_bias.pgf}
    \caption{$\log_{10}(M/h^{-1}M_{\odot}) = 13.0-13.5$}  
    \vspace{4ex}
  \end{subfigure} 
  \begin{subfigure}[b]{0.49\textwidth}
    \centering
    \input{EFT_likelihood/13.5-14/13.5-14.0_different_bias.pgf} 
    \caption{$\log_{10}(M/h^{-1}M_{\odot}) = 13.5-14.0$} 
  \end{subfigure}
  \begin{subfigure}[b]{0.49\textwidth}
    \centering
    \input{EFT_likelihood/14-14.5/14.0-14.5_different_bias.pgf} 
    \caption{$\log_{10}(M/h^{-1}M_{\odot}) = 14.0-14.5$} 
  \end{subfigure} 
  \caption{MAP value $\hat{\beta}$ found using the EFT likelihood for two bias orders at $z = 0$. The different sub-figures show four different mass ranges.}
  \label{fig: z=0 EFT}
\end{figure}

In this section, we show the results of applying the EFT likelihood to the halo catalogs. We start by comparing the results for two different bias orders---second and third order---at fixed redshift $z=0$. 
Fig. \ref{fig: z=0 EFT} shows the deviation of the MAP values $\hat{\beta}$ from 1 as a function of $\Lambda$ for different halo mass ranges. For all of halo mass bins except the highest one, $\hat{\beta}$ is consistent with being unbiased within the error bar obtained from the profile likelihood. Moreover, $\hat{\beta}$ is moving closer to 1 as $\Lambda$ is increased, consistent with the shrinking error bar as more $k$ modes are being included in the likelihood and forward model. 
We notice that the MAP values $\hat{\beta}$ are closer to 1 for the third order bias expansion than in the case of second order, for every halo sample. This indicates that the systematic error in $\hat{\beta}$ in the 3rd order bias case is reduced, as expected if one is in the converging regime of the EFT. Therefore, in the rest of the paper, we focus only on the 3rd order bias expansion.

\begin{figure}[h]
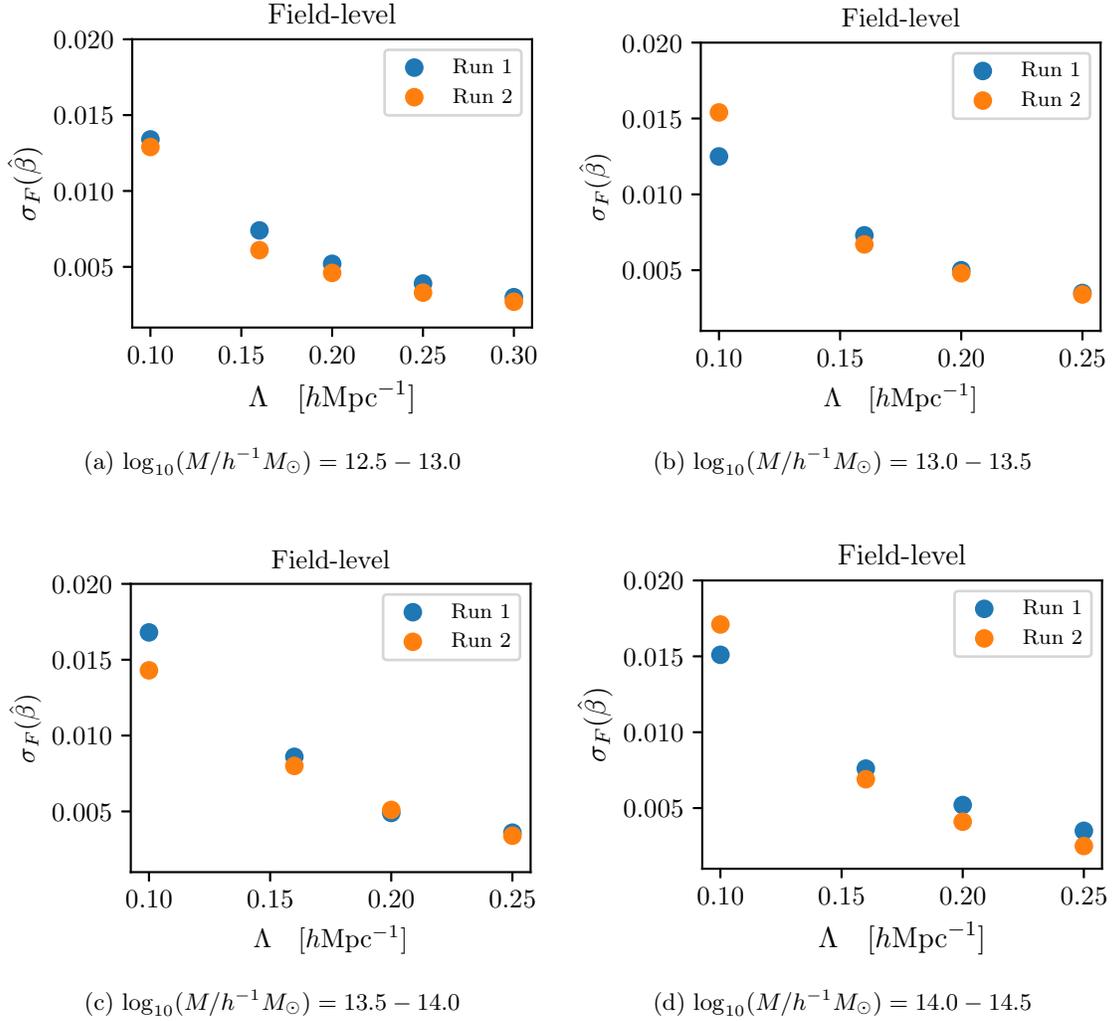

  \begin{subfigure}[b]{0.49\textwidth}
    \centering
    \input{EFT_likelihood/12.5-13/12-13_z=0_error.pgf} 
    \caption{$\log_{10}(M/h^{-1}M_{\odot}) = 12.5-13.0$} 
    \vspace{4ex}
  \end{subfigure}
  \begin{subfigure}[b]{0.49\textwidth}
    \centering
   \input{EFT_likelihood/13_13.5/13-13.5_z=0_error.pgf} 
    \caption{$\log_{10}(M/h^{-1}M_{\odot}) = 13.0-13.5$}  
    \vspace{4ex}
  \end{subfigure} 
  \begin{subfigure}[b]{0.49\textwidth}
    \centering
    \input{EFT_likelihood/13.5-14/A13.5-14_z=0_error.pgf} 
    \caption{$\log_{10}(M/h^{-1}M_{\odot}) = 13.5-14.0$} 
  \end{subfigure}
  \begin{subfigure}[b]{0.49\textwidth}
    \centering
    \input{EFT_likelihood/14-14.5/14-14.5_z=0_error.pgf} 
    \caption{$\log_{10}(M/h^{-1}M_{\odot}) = 14.0-14.5$} 
  \end{subfigure} 
  \caption{$\sigma_F(\hat{\beta)}$ values as a function of $\Lambda$ at $z=0$. Different sub-figures show four different mass ranges.}
  \label{fig: EFT error}
\end{figure}

Fig. \ref{fig: EFT error} depicts the value of the 1$\sigma$ error bar, $\sigma_F(\hat{\beta)}$, for the field-level inference (as emphasized by the subscript $F$) as a function of $\Lambda$ for the 3rd bias order for both runs 1 and 2. We see that $\sigma_F(\hat{\beta)}$ is smoothly decreasing with increasing $\Lambda$. Since our initial conditions are exactly the ones used in the halo simulations, the statistical uncertainty $\sigma_F(\hat{\beta)}$ is only sourced by the halo stochasticity which appears in the EFT likelihood. 
Note that we do expect the $\sigma_F(\hat{\beta)}$ results to change once we start sampling the initial phases instead of keeping them fixed.

We also notice that the $\sigma_F(\hat{\beta)}$ values do not change much between the different halo samples. This trend can be understood by inspecting how the numerator and denominator of Eq. \eqref{eq: EFT likelihood} change with halo mass. On the one hand, more massive halos are rarer, and hence have larger noise (stochasticity), i.e. larger $\sigma_{\varepsilon}^{2}$ in the denominator. On the other hand, higher-mass halos are more biased, and hence show a stronger clustering signal.
Hence the numerator also increases with halo mass. As a consequence, the ratio of both quantities is actually roughly constant, so that we get a similar $\sigma_F(\hat{\beta)}$ for all halo bins considered. Notice that this result only applies to the \emph{fixed-phase} study done here.

\begin{figure}[h]
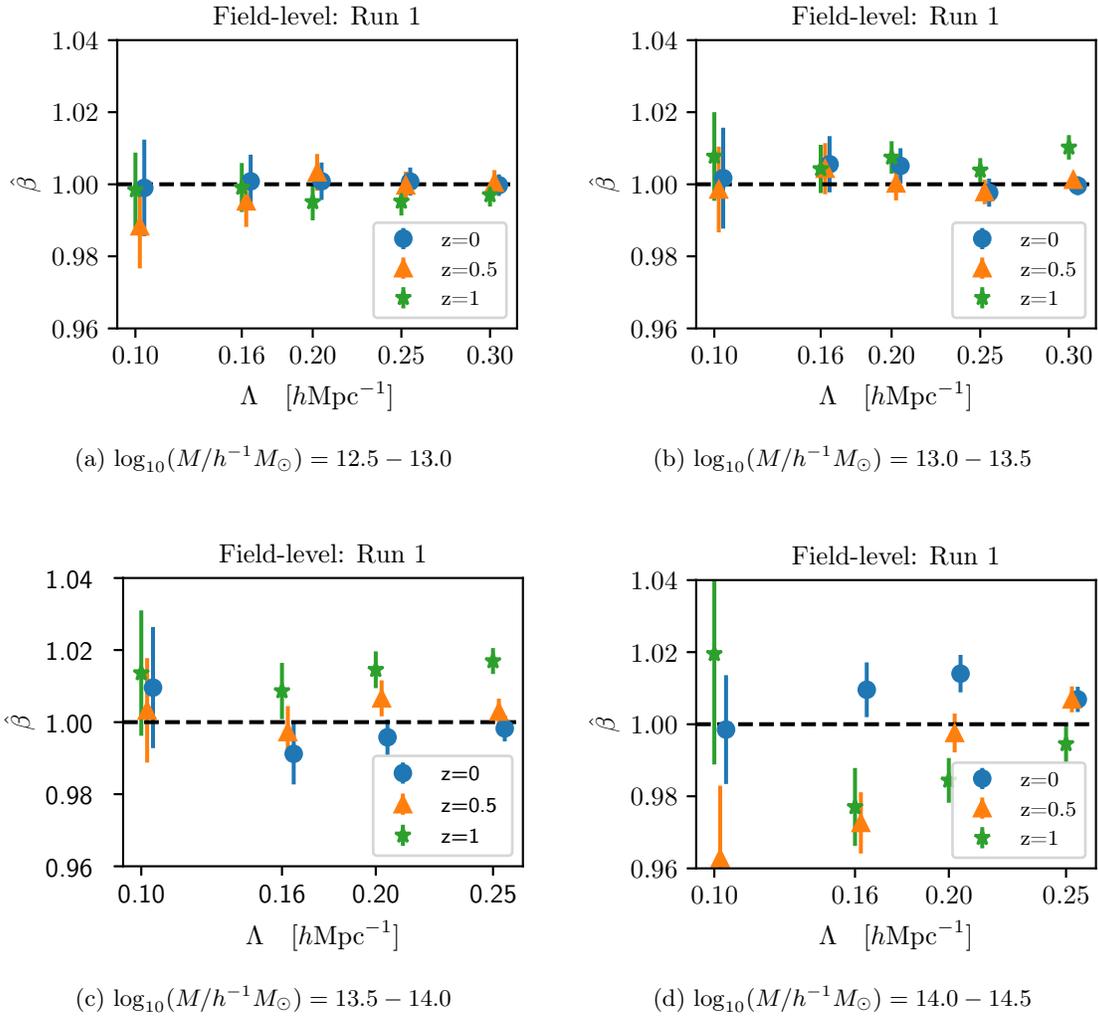

\centering
  \begin{subfigure}[b]{0.49\textwidth}
    \centering
    \input{EFT_likelihood/12.5-13/12.5-13_different_z.pgf} 
    \caption{$\log_{10}(M/h^{-1}M_{\odot}) = 12.5-13.0$} 
    \vspace{4ex}
  \end{subfigure}
  \begin{subfigure}[b]{0.49\textwidth}
    \centering
    \input{EFT_likelihood/13_13.5/13-13.5_different_z.pgf} 
    \caption{$\log_{10}(M/h^{-1}M_{\odot}) = 13.0-13.5$}  
    \vspace{4ex}
  \end{subfigure} 
  \begin{subfigure}[b]{0.49\textwidth}
    \centering
    \input{EFT_likelihood/13.5-14/13.5-14_different_z.pgf} 
    \caption{$\log_{10}(M/h^{-1}M_{\odot}) = 13.5-14.0$} 
  \end{subfigure}
  \begin{subfigure}[b]{0.49\textwidth}
    \centering
    \input{EFT_likelihood/14-14.5/14-14.5_different_z.pgf} 
    \caption{$\log_{10}(M/h^{-1}M_{\odot}) = 14.0-14.5$} 
  \end{subfigure} 
  \caption{MAP values for $\beta$ using the EFT likelihood found at different redshifts for run 1. Different panels show four different mass ranges at three different redshifts each.
  }
  \label{fig: EFT different mass bins}
\end{figure}

Next, we look at the results found at different redshifts. Fig. \ref{fig: EFT different mass bins} and Fig. \ref{fig:Run2 EFT different mass bins} show the value of $\hat{\beta}$ as a function of $\Lambda$ at redshifts $z = 0.0, 0.5, 1.0$ for run 1 and run 2, respectively. For run 1, these results are also summarized in  Tab. \ref{table: bias3Allz} for a fixed cutoff,  $\Lambda = 0.16 \invMpc$. We find that the remaining systematic bias is very low across all the redshifts and mass ranges. Even for this cutoff, the bias is less than 1\% at $z=0$ for all mass ranges and the results generally keep improving with growing $\Lambda$. If we look across all redshifts, it is clear that the remaining bias is still below 2\%, and in fact consistent with zero, for most of the cases. It goes over 2\% only for the most massive halos which are very rare at higher redshifts. Those halo samples would most likely benefit from from going to higher bias orders in the bias expansion.

From  Fig. \ref{fig: EFT different mass bins} and Fig. \ref{fig:Run2 EFT different mass bins} we notice that the remaining systematic bias in $\hat{\beta}$ is increasing with growing redshift. This occurrence is counter-intuitive, since from perturbation theory we would expect a better performance at higher redshifts where  perturbation theory extends to higher wavenumbers. A similar trend was noticed with inference of $\sigma_{8}$ from the halo catalogues described in \cite{Schmidt_sigma8}. It was found there that this trend is caused by the higher-order bias terms. Although the higher-order bias terms are suppressed by powers of the normalized growth factor $D_{\rm {norm}}(z) = D(z)/D(0)$ at higher redshifts, it is possible that the increase in their coefficients with redshift more than compensates for this suppression.
To check if this was the case for us as well, we use the test suggested in \cite{Schmidt_sigma8} which was based on the assumption from \cite{large_scale_bias_2018} that the higher order bias terms can be approximated as being a function of $(b_{1}-1)D_{\rm {norm}}(z)$. Results are shown in Fig. \ref{fig: frac_err}, where we plotted the $|\hat{\beta}-1|$ values for all halo mass bins and redshifts against $(b_{1}-1)D_{\rm {norm}}(z)$. There is a hint of a correlation between $|\hat{\beta}-1|$ and $(b_{1}-1)D_{\rm {norm}}(z)$, although all but one points are consistent with $\hat{\beta}=1$ within one sigma.

Let us also comment on the limits of the cutoff we are using. For matter, the EFT is under perturbative control for $\Lambda \lesssim 0.25 \invMpc$ at $z=0$. For highly biased tracers, the cutoff is reduced due to the growing size of bias parameters at higher orders. Thus, we are going beyond that limit, and not all of our values of $\Lambda$ are strictly under perturbative control. 
However, because the BAO is an oscillatory feature, while higher-order corrections are expected to be smooth functions of $k$, the BAO inference seems to be still robust at these high $k$. We leave a more systematic investigation of this important issue to future work. 

\begin{figure}[h]
\centering
  \begin{subfigure}[b]{0.49\textwidth}
    \centering
    \input{EFT_likelihood/2nd_Run/12.5-13_different_z_Run2.pgf} 
    \caption{$\log_{10}(M/h^{-1}M_{\odot}) = 12.5-13.0$} 
    \vspace{4ex}
  \end{subfigure}
  \begin{subfigure}[b]{0.49\textwidth}
    \centering
    \input{EFT_likelihood/2nd_Run/13.0-13.5_different_z_Run2.pgf} 
    \caption{$\log_{10}(M/h^{-1}M_{\odot}) = 13.0-13.5$}  
    \vspace{4ex}
  \end{subfigure} 
  \begin{subfigure}[b]{0.49\textwidth}
    \centering
    \input{EFT_likelihood/2nd_Run/13.5-14.0_different_z_Run2.pgf} 
    \caption{$\log_{10}(M/h^{-1}M_{\odot}) = 13.5-14.0$} 
  \end{subfigure}
  \begin{subfigure}[b]{0.49\textwidth}
    \centering
    \input{EFT_likelihood/2nd_Run/14.0-14.5_different_z_Run2.pgf} 
    \caption{$\log_{10}(M/h^{-1}M_{\odot}) = 14.0-14.5$} 
  \end{subfigure} 
  \caption{MAP values for $\beta$ using the EFT likelihood found at different redshifts for run 2. Different panels show four different mass ranges at three different redshifts each. 
  }
  \label{fig:Run2 EFT different mass bins}
\end{figure}

\begin{figure}[h!]
    \centering
    \input{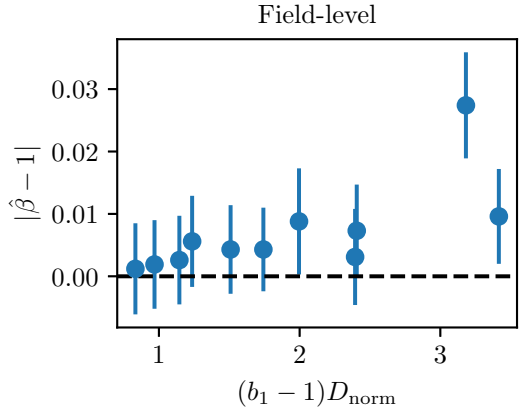}
    \caption{MAP values for $|\hat\beta - 1|$ found at the cutoff $\Lambda = 0.16 \invMpc$ for all halo mass bins and redshifts against $(b_{1}-1)D_{\rm {norm}}(z)$, where $D_{\rm {norm}}(z) = D(z)/D(0)$.}\label{fig: frac_err}
\end{figure}

\begin{table}
\centering
\begin{tabular}{c c c c} 
 \hline
 \hline
  $z$ & Mass range $\log_{10}(M/h^{-1}M_{\odot})$ & $100(\hat{\beta}-1)$ & $\sigma_{\varepsilon}$  \\ 
 \hline
 \hline
 0.0 & $[12.5-13.0]$ & 1.08 $\pm$ 0.74 & 0.463\\ 
 0.5 & $[12.5-13.0]$ & -0.47 $\pm$ 0.71 &  0.471\\ 
 1.0 & $[12.5-13.0]$ & -0.09 $\pm$ 0.68 & 
 0.494\\
 \hline
 0.0 & $[13.0-13.5]$ & 0.56 $\pm$ 0.78 & 0.625\\
 0.5 & $[13.0-13.5]$ & 0.40 $\pm$ 0.70 &  0.647 \\
 1.0 & $[13.0-13.5]$ & 0.43 $\pm$ 0.67 & 0.735\\
  \hline
 0.0 & $[13.5-14.0]$ & -0.89 $\pm$ 0.86 & 0.992\\
 0.5 & $[13.5-14.0]$ & -0.28 $\pm$ 0.72 & 1.163\\
 1.0 & $[13.5-14.0]$ & 0.86 $\pm$ 0.75 & 1.582\\
 \hline
 0.0 & $[14.0-14.5]$ & 0.80 $\pm$ 0.80 & 1.785\\
 0.5 & $[14.0-14.5]$ & -2.73 $\pm$ 0.80 & 2.572\\
 1.0 & $[14.0-14.5]$ & -2.29 $\pm$ 0.97 & 4.926\\ 
 \hline
 \hline
\end{tabular}
\caption{Summary of the results found using the field-level EFT likelihood at the cutoff $\Lambda=0.16 \invMpc $ for different redshifts and halo mass bins.  }\label{table: bias3Allz}
\end{table}

\clearpage
\newpage

\section{Comparing the field-level results to the power spectrum approach}\label{sec: PS results}

Having presented the results of constraining the BAO scale using the EFT likelihood, we now turn to comparing these results to a more traditional BAO inference approach based on the power spectrum.

\subsection{Power spectrum likelihood}

Care is needed in order to ensure that the comparison we are making is valid, since in the EFT approach we use fixed phases in the matter density field. 
Therefore, we adopt the following Gaussian likelihood for the halo power spectrum:
\begin{equation}\label{eq: Gauss likelihood}
    -2\ln \mathcal{L}[P_{h}(k) | \delta_{\rm in}, \{b_O\}, P_\varepsilon] = \sum_{k}^{k_{\rm max}} 
    \frac{[P_{h}(k)-P_{\varepsilon} - P_{\rm det}(k|\delta_{\rm in}, \beta, \{b_{O}\}
    ]^{2}}{{\rm Var}_{{\rm fix}} [P_h(k)]}.
\end{equation}

Here, $P_{\rm det}(k|\delta_{\rm in}, \beta, \{b_{O}\})$ is the power spectrum of the deterministic halo field found using the same forward model as in EFT case for a fixed $\beta$ value; $P_{h}(k)$ is the measured halo power spectrum, $m_{k}$ is the number of modes in a wavenumber bin, and $P_{\varepsilon}$ is the noise spectrum. 
Notice that the covariance appearing in the numerator of the likelihood, ${\rm Var}_{{\rm fix}} [P_h(k)]$, is modified to reflect the fact that we are using fixed phases. The derivation of the power spectrum covariance for fixed phases can be found in Appendix \ref{sec: Variance} and its final form is given in Eq. \eqref{eq: prediction variance}. It is also important to note that we are not performing any additional BAO reconstruction on the halo data, but comparing the halo power spectrum directly with the theory predictions from the full forward model. Therefore the comparison we are making is at the level of likelihoods: the EFT likelihood is performing at the level of the field, while the likelihood in Eq. \eqref{eq: Gauss likelihood} compresses the data to the power spectrum in bins of $k$. Both likelihoods however consistently assume fixed initial conditions.

To find the best fit for $\beta$, we use the following procedure for different values $\beta_{i}$. We start by finding the initial matter fields with the BAO scales $r_{s} = \beta_{i} r_{\rm fid}$ using Eq. \eqref{eq: denisty} as in the field-level likelihood calculations. 
Once we have the linear matter density field $\delta_{\rm in}(k, \beta_i)$ , we use the 3LPT forward model to generate the evolved matter field $\delta(k,\beta)=\delta_{\rm fwd}[\delta_{\rm in}(k, \beta_{i})]$, where we set all modes with $k>\Lambda$
to zero. 
For the bias operators, we use the same bias model as described in Sec. \ref{About eft}.
The MAP for the bias parameters is found by maximizing the EFT likelihood. We keep one bias parameter free at a time  and marginalize over all other bias coefficients. Once we found the MAP value for that parameter, we  move on and repeat the procedure for the remaining ones.  This gives us the deterministic halo field whose power spectrum  $P_{\rm det}(k|\delta_{\rm in}, \beta, \{b_{O}\})$ is straightforward to measure in the same $k$ bins as the halo sample.

We now turn to the determination of $P_\varepsilon$. Ideally, one would fit for this together with $\beta$ and the bias parameters. In our simplified analysis, we only fit $P_\varepsilon$, and use the same noise spectrum value $P_{\varepsilon}$ across all $\Lambda$ and $\beta$ values. This value is found for $\Lambda = 0.2$ $h \rm{Mpc}^{-1}$ and $\beta = 1.00$ by fitting the difference $P_h(k) - P_{\rm det}(k|\delta_{\rm in}, \beta=1, \{b_{O}\})$  to a constant, using $w = 1/\sigma_{\rm w}$ as the weight where $\sigma_{\rm w} = |P_{h} - P_{\rm det}|/\sqrt{2/m_k}$. Fitting the noise separately from bias terms and $\beta$ leaves us with some uncertainties in its estimate. We roughly estimate this uncertainty by repeating the same analysis for run 2 halo samples, resulting in values of $P_{\varepsilon}$ that differ by around 20\%, which results in a corresponding 20\% shift in the $1\sigma$ error for $\hat{\beta}$. We conclude that our results for the latter carry an uncertainty of $\sim 20\%$. This is sufficient for the approximate comparison we are aiming for in this paper. We aim to improve this in future work.

\begin{figure}[h!]
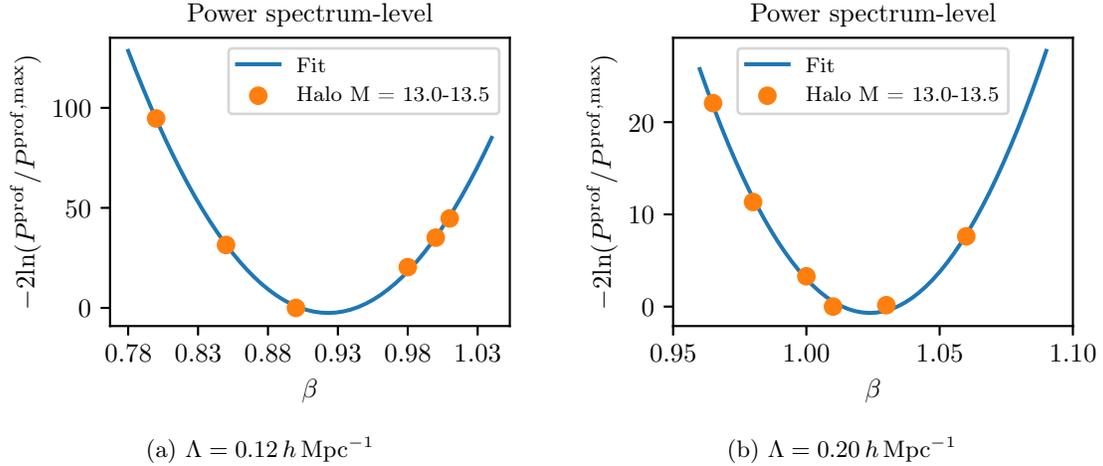

\centering
\begin{subfigure}{.5\textwidth}
  \centering
  \input{Gauss_likelihood/Parabola Fit/Standard_Likelihood_0.12.pgf}
  \caption{$\Lambda = 0.12 \invMpc$}
\end{subfigure}%
\begin{subfigure}{.5\textwidth}
  \centering
  \input{Gauss_likelihood/Parabola Fit/Standard_Likelihood_beta_lambda=0.2.pgf}
  \caption{$\Lambda = 0.20 \invMpc$
  }
\end{subfigure}
\caption{Profile likelihood $-2\ln P^{\rm prof}$ for the power spectrum, plotted as a function of $\beta$ for two different cutoffs $\Lambda$ at $z=0$. The blue line shows the parabolic fit which was used to find MAP $\hat{\beta}$ and $\sigma_{PS}(\hat{\beta})$ error.}\label{fig. parabola}
\label{fig:test}
\end{figure}

Finally, by inserting $P_{\rm det}(k|\delta_{\rm in}, \beta_i, \{b_{O}\})$  and $P_{\varepsilon}$ in Eq. \eqref{eq: Gauss likelihood}, we find the likelihood value for each $\beta_{i}$. 
Repeating this procedure at fixed halo sample, redshift and $\Lambda$, leads to a set $\{\beta^{i}, -2\ln P^{\rm prof}(\beta^{i})  \}$, which is nicely fit by a parabola. An example of this parabola fit is shown in  Fig. \ref{fig. parabola}.
$\hat{\beta}$ and $\sigma_{PS}(\hat{\beta})$, the value of the $1\sigma$ error bar for the power spectrum inference, are found as the location of the minimum and the inverse square root of the parabolic fit, respectively.

\subsection{Results}\label{sec: Ps plots}

\begin{table}
\centering
\begin{tabular}{  c c c c  } 
  z & Mass range $\log_{10}(M/h^{-1}M_{\odot})$ & $100(\hat{\beta}-1)$ & $b_{1}$  \\ 
 \hline
 \hline
 0.0 & $[12.5-13.0]$ & 1.20 $\pm$ 0.90 & 0.833\\ 
 0.5 & $[12.5-13.0]$ & 1.87 $\pm$ 0.79 & 1.266\\ 
 1.0 & $[12.5-13.0]$ & 1.32 $\pm$ 0.73 & 1.901\\
 \hline
 0.0 & $[13.0-13.5]$ & 4.80 $\pm$ 0.96 & 1.236\\
 0.5 & $[13.0-13.5]$ & 0.79 $\pm$ 0.94 & 1.973\\
 1.0 & $[13.0-13.5]$ & -2.01 $\pm$ 0.81 & 2.892\\
  \hline
 0.0 & $[13.5-14.0]$ & 2.27 $\pm$ 1.27 & 1.996\\
 0.5 & $[13.5-14.0]$ & 1.77 $\pm$ 1.01 & 3.129\\
 1.0 & $[13.5-14.0]$ & 1.41 $\pm$ 0.88 & 3.994\\
 \hline
 0.0 & $[14.0-14.5]$ & 1.18 $\pm$ 1.46 & 3.416\\
 \hline
 \hline
\end{tabular}
\caption{MAP values of $\beta$ for cutoff $\Lambda=0.16 \invMpc$ inferred from the power spectrum likelihood, at different redshifts for different halo mass ranges.}\label{Table: Gauss}
\end{table}

\begin{figure}[h!]
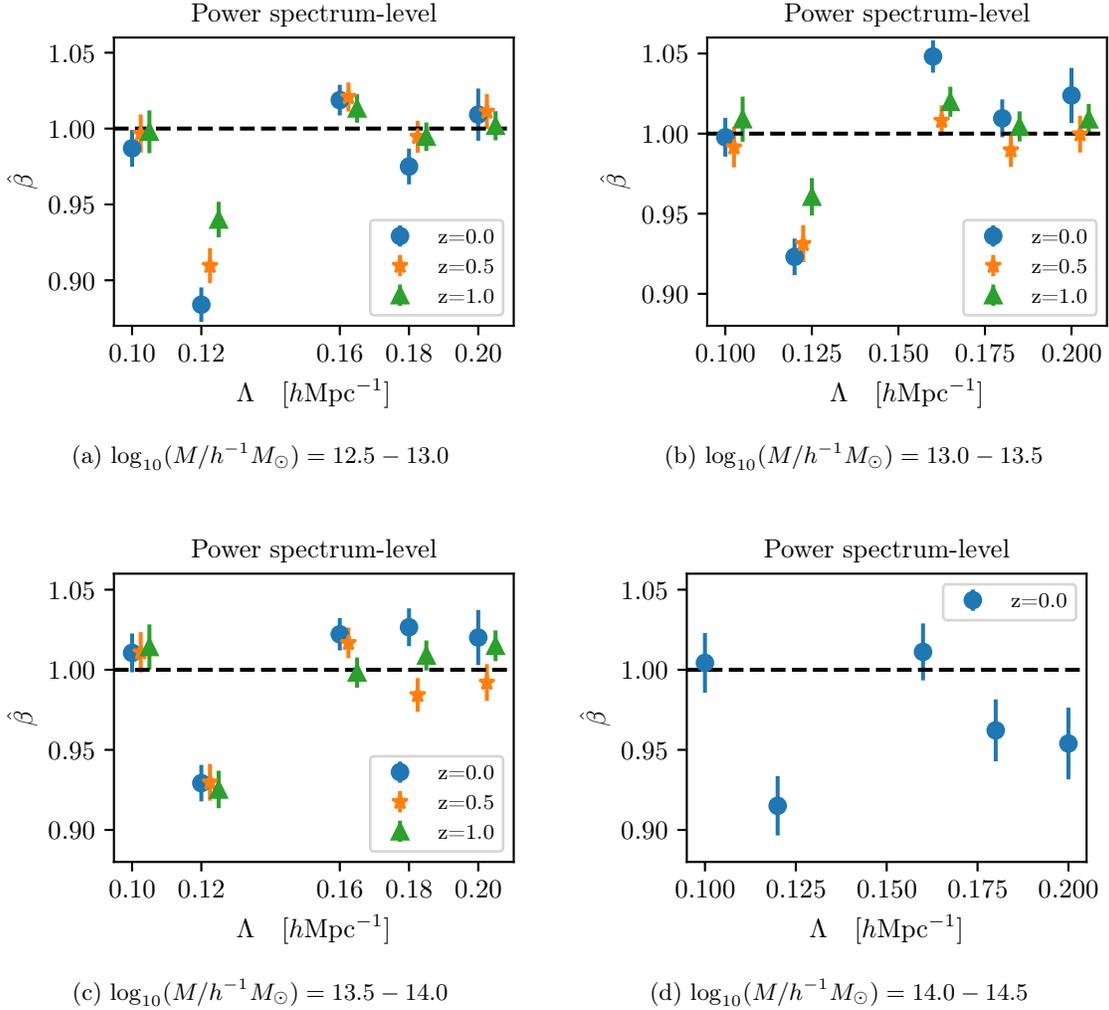

  \begin{subfigure}[b]{0.49\textwidth}
    \centering
    \input{Gauss_likelihood/All_redshifts/12.5-13.0_different_z.pgf}
    \caption{$\log_{10}(M/h^{-1}M_{\odot})= 12.5-13.0$} 
    \vspace{4ex}
  \end{subfigure}
  \begin{subfigure}[b]{0.49\textwidth}
    \centering
    \input{Gauss_likelihood/All_redshifts/13.0-13.5_different_z.pgf} 
    \caption{$\log_{10}(M/h^{-1}M_{\odot})= 13.0-13.5$}  
    \vspace{4ex}
  \end{subfigure} 
  \begin{subfigure}[b]{0.49\textwidth}
    \centering
    \input{Gauss_likelihood/All_redshifts/13.5-14.0_different_z.pgf} 
    \caption{$\log_{10}(M/h^{-1}M_{\odot})= 13.5-14.0$} 
  \end{subfigure}
  \begin{subfigure}[b]{0.49\textwidth}
    \centering
    \input{Gauss_likelihood/All_redshifts/14.0-14.5_different_z.pgf} 
    \caption{$\log_{10}(M/h^{-1}M_{\odot})= 14.0-14.5$} 
  \end{subfigure} 
  \caption{MAP values for $\beta$ using the power spectrum likelihood for different $\Lambda$. Different panels show four different mass ranges at three different redshifts. }
  \label{fig: PS different mass bins}
\end{figure}

We now turn to the results for MAP. $\hat{\beta}$ is found using the likelihood given in Eq. \eqref{eq: Gauss likelihood}.
The residual values of $\hat{\beta}$ as a function of $\Lambda$ at the three different redshifts are shown in Fig. \ref{fig: PS different mass bins}. For the most massive halo range $\log_{10}(M/h^{-1}M_{\odot}) = 14.0-14.5$, we show results only at redshift zero. For this halo range at higher redshifts, the set  $\{\beta^{i}, -2\ln P^{\rm prof}(\beta^{i})  \}$ does not yield a well-defined maximum. We also exclude all the samples for which the MINUIT algorithm does not converge  for the bias coefficients due to a  poor signal to noise ratio.

\begin{figure}[h!]
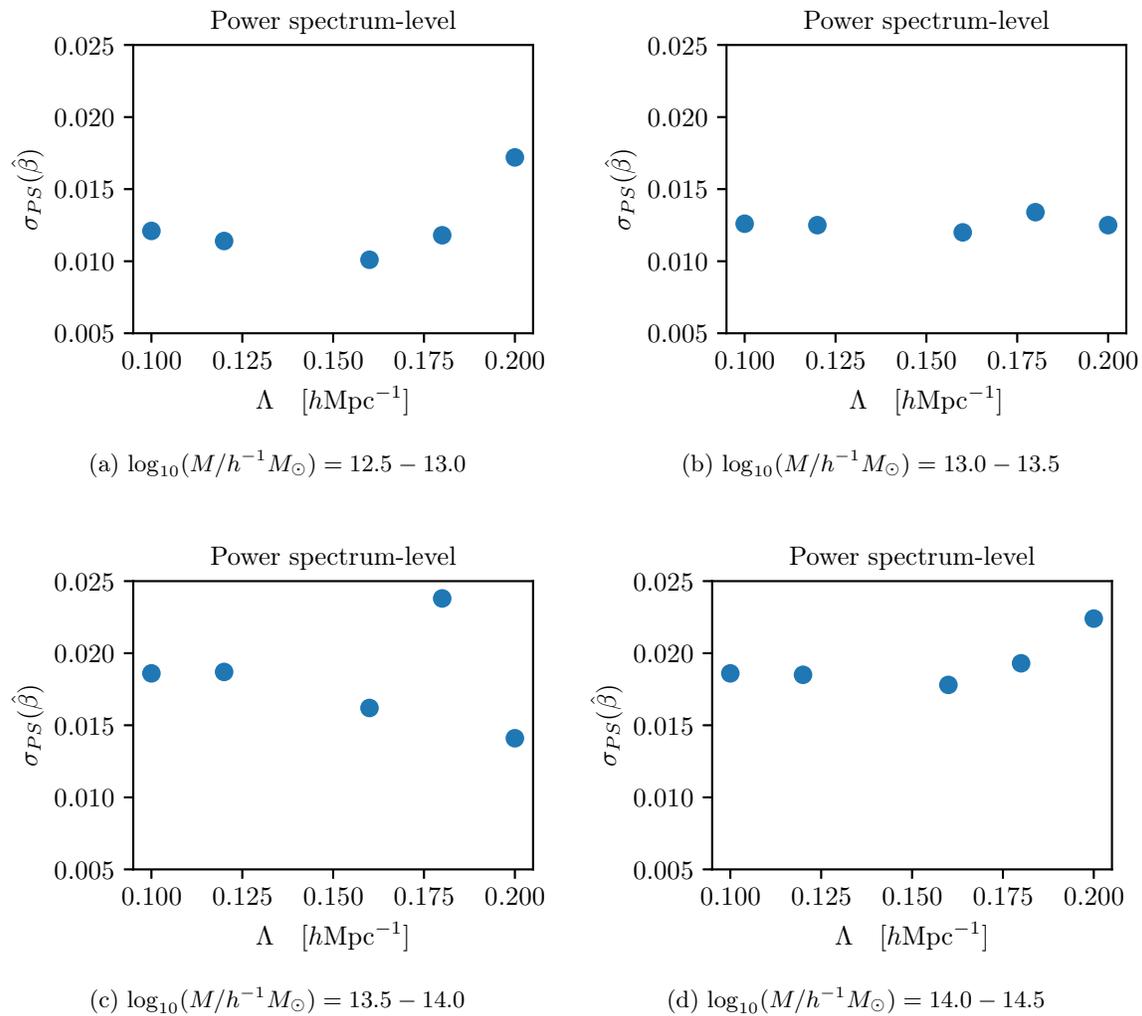

  \begin{subfigure}[b]{0.49\textwidth}
    \centering
    \input{Gauss_likelihood/12.5-13.0/12error_plot_vs_Lambda.pgf}
    \caption{$\log_{10}(M/h^{-1}M_{\odot})= 12.5-13.0$} 
    \vspace{4ex}
  \end{subfigure}
  \begin{subfigure}[b]{0.49\textwidth}
    \centering
    \input{Gauss_likelihood/13.0-13.5/13error_plot_vs_Lambda.pgf} 
    \caption{$\log_{10}(M/h^{-1}M_{\odot})= 13.0-13.5$}  
    \vspace{4ex}
  \end{subfigure} 
  \begin{subfigure}[b]{0.49\textwidth}
    \centering
    \input{Gauss_likelihood/13.5-14.0/13.5error_plot_vs_Lambda.pgf} 
    \caption{$\log_{10}(M/h^{-1}M_{\odot})= 13.5-14.0$} 
  \end{subfigure}
  \begin{subfigure}[b]{0.49\textwidth}
    \centering
    \input{Gauss_likelihood/14.0-14.5/14error_plot_vs_Lambda.pgf} 
    \caption{$\log_{10}(M/h^{-1}M_{\odot})= 14.0-14.5$} 
  \end{subfigure} 
  \caption{$\sigma_{PS}(\hat{\beta})$ values found using the power spectrum likelihood for different $\Lambda$. Different panels show four different mass ranges at the redshift $z=0$.}
  \label{fig: sigma different mass bins}
\end{figure}

The quantitative results are summarized in Tab. \ref{Table: Gauss}. We see that, for most of the samples, the residual bias in $\hat{\beta}$ is between 1.20\% and 2.3\%. The MAP values of the linear bias parameter $b_1$ are also listed in the table. We notice that, for a fixed mass bin, $b_1$ is increasing with halo mass and redshift as is expected. 
In  Fig. \ref{fig: sigma different mass bins}, we show  $\sigma_{PS}(\hat{\beta)}$ as a function of $\Lambda$ at redshift $z=0$. While for the field-level likelihood $\sigma_{F}(\hat{\beta)}$ reduces about 2.4 times from $\Lambda = 0.1 \invMpc$ to $\Lambda = 0.2 \invMpc$, here we do not see such a trend. Instead, $\sigma_{PS}(\hat{\beta)}$ stays fairly constant across all $\Lambda$ for the power spectrum likelihood. This is presumably because the field-level likelihood can still make use of the phase information at wavenumbers for which the power spectrum likelihood is already dominated by the noise $P_\varepsilon$.\\

\begin{figure}
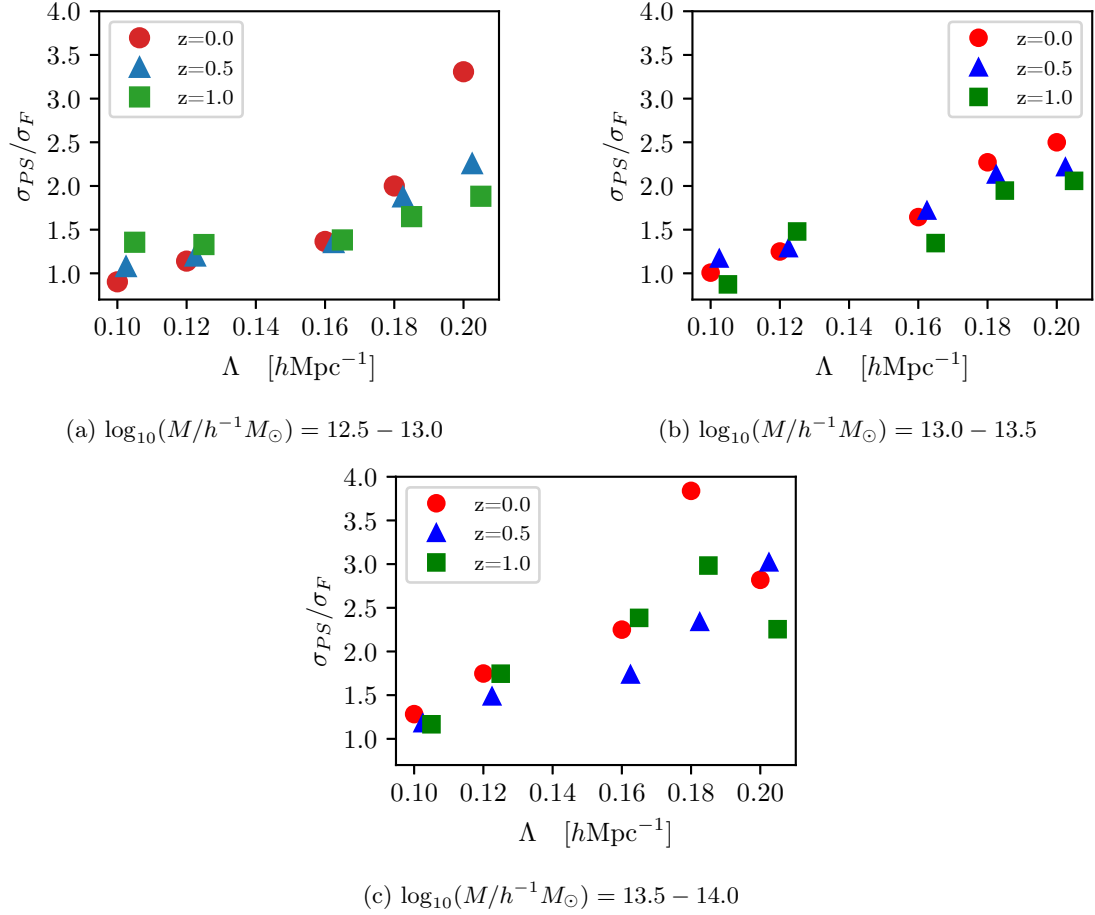

     \centering
     \begin{subfigure}[b]{0.49\textwidth}
         \centering
         \input{Gauss_likelihood/12.5-13.0/different_z_M12_error_bar_ratio.pgf}
    \caption{$\log_{10}(M/h^{-1}M_{\odot})= 12.5-13.0$} 
         \label{fig:y equals x}
     \end{subfigure}
     \hfill
     \begin{subfigure}[b]{0.49\textwidth}
         \centering
         \input{Gauss_likelihood/13.0-13.5/different_z_M13_error_bar_ratio.pgf}
    \caption{$\log_{10}(M/h^{-1}M_{\odot})= 13.0-13.5$}
     \end{subfigure}
     
     \begin{subfigure}[b]{0.49\textwidth}
         \centering
         \input{Gauss_likelihood/13.5-14.0/different_z_M13_5_error_bar_ratio.pgf} 
    \caption{$\log_{10}(M/h^{-1}M_{\odot})= 13.5-14.0$}
         \label{fig:five over x}
     \end{subfigure}
        
  \caption{Ratio of the uncertainty on the BAO scale inferred from the power spectrum likelihood, $\sigma_{PS}(\hat{\beta})$, to that from the field-level likelihood, $\sigma_{F}(\hat{\beta})$, as a function of cutoff for different redshifts. Each panel corresponds to a different halo mass range.}\label{Fig: ratio 13}
\end{figure}

The most interesting result is Fig. \ref{Fig: ratio 13}, 
which compares the error on $\hat{\beta}$ from the power spectrum approach, 
$\sigma_{PS}(\hat{\beta})$, to the one from the field level approach, $\sigma_{F}(\hat{\beta})$. This ratio is shown for three different halos mass ranges at three different redshifts. For smaller cutoffs, both likelihoods give similar results, which is the expected result if the data ($\delta_h$) are well approximated as a Gaussian random field. However, as $\Lambda$ grows, the EFT likelihood starts outperforming the power spectrum based likelihood. At the highest $\Lambda$ considered, the $\sigma_{F}(\hat{\beta})$ value is around 2.5 times smaller than $\sigma_{PS}(\hat{\beta})$. The field-level EFT likelihood performs better because it operates at the level of the field. This means that it includes not only all the information coming from the power spectrum, but also information from the from N-point functions of arbitrarily high orders. Concretely in the case of the BAO, the field-level likelihood knows about the bulk flow field, and can thus compare the expected BAO scale at a given location with the data. The power spectrum on the other hand is averaged over all locations, and thus suffers from the damping of the BAO peak \cite{padmanabhan/etal:2009,sherwin/zaldarriaga}.
Thus, the fact that the field-level likelihood outperforms the power spectrum based one comes as no surprise.

Finally, note that we have fixed the bias coefficients in the theory prediction for the power spectrum to the values obtained from the field-level likelihood. In practice, those would have to be marginalized over in a power spectrum analysis.

\clearpage
\newpage
\section{Summary and Conclusions}
\label{sec: Conclusion}

In this paper we compared the inference of the BAO scale from the halo catalogs using an LPT-based forward model combined with the EFT likelihood with the standard approach which compresses the data to the power spectrum. The forward model uses a combination of 3LPT expansion for the matter field and a third-order bias expansion. Our results were expressed in the terms of the parameter $\beta$, defined as the ratio of the measured value of the BAO scale to its fiducial value. 

The field-level inference results are summarised in Fig. \ref{fig: EFT different mass bins} and Tab. \ref{table: bias3Allz}. From these it is clear that the remaining systematic error in $\hat{\beta}$ is at most $\sim 2$\% for all samples. 
If we ignore the most biased sample ($\log_{10}(M/h^{-1}M_{\odot}) > 14.0$ at $z=0.5$ and $z=1.0$), at $\Lambda=0.16\invMpc$, the remaining bias in $\beta$ is less than 1\%  for all remaining samples, which is remarkably low. Moreover, $\hat{\beta}$ is statistically consistent with being unbiased for all halo samples except the highest mass one. It is also interesting to notice that the bias in $\hat{\beta}$ is under control for all halo mass ranges, even for $\Lambda = 0.25 \invMpc$, which is close to the nonlinear scale. For the lighter halos, $\log_{10}(M/h^{-1}M_{\odot}) < 13.5$, this even applies to $\Lambda = 0.3 \invMpc$.

While we consider halos here, the EFT approach is equally applicable to galaxies. This is confirmed by the results of \cite{barreira/etal:2021}, who demonstrated an unbiased inference of the linear power spectrum normalization $\sigma_8$ on fully hydrodynamical simulated galaxies.

In order to assess the performance of the field-level inference of the BAO scale, we compare it to the more traditional approach of the BAO inference from the power spectrum. For this we utilized the likelihood defined in Eq. \eqref{eq: Gauss likelihood}, where the theory model for the power spectrum is based on the same forward model as in the EFT likelihood approach (in particular, the field-level likelihood was still used to find the best-fit values for the bias parameters). We modified the covariance in this likelihood to reflect the fact that we are using fixed phases, so that we can compare the two approaches on the same footing. Results found using this likelihood are shown in Fig. \ref{fig: PS different mass bins} and Tab. \ref{Table: Gauss}. For a fixed $\Lambda = 0.16 \invMpc$, the remaining systematic error $\hat{\beta}$ is between  1.2\% and 2.3\% for those halo samples that yielded converged profile likelihoods. 

Fig. \ref{Fig: ratio 13} shows the relative performance of the field-level and power spectrum based likelihoods. Across all halos samples, we notice a similar trend. For smaller cutoffs, both likelihoods show similar performance. However, for $\Lambda>0.12 \invMpc$, the field-level likelihood gives better results across all the halo masses and redshifts.  
For the highest cutoff value considered in both likelihoods, $\Lambda = 0.2\invMpc$, the error on the BAO scale inferred from the power spectrum is between $2.47-3.3$ times larger than that obtained from the field-level likelihood, depending on the halo sample. Since the field-level likelihood contains all the information that would come from the higher order correlation functions, including the precise bulk-flow field, while the information available in the likelihood from Eq. \eqref{eq: Gauss likelihood} are only those from the power spectrum, a better performance of the EFT likelihood was to be expected. 

In future work we will investigate how well we can constrain the BAO from the EFT likelihood in the cases when the initial conditions are not fixed, but sampled. This will allow for a realistic comparison of the constraining power on the BAO scale that can be obtained from the field-level inference as compared to that based on the galaxy power spectrum.

\acknowledgments
We would like to thank Rodrigo Voivodic and Laura Herold for helpful discussions.
IB would like to thank to  BAYHOST Scholarship programs sponsored by the Free State of Bavaria for graduates of Central, Eastern and Southeastern European states.
FS acknowledges support from the Starting Grant (ERC-2015-STG 678652) ``GrInflaGal'' of the European Research Council.

\clearpage
\appendix
\section{Power spectrum covariance for fixed phases}\label{sec: Variance}

\begin{figure}[h!]
    \centering
    \scalebox{0.5}{\input{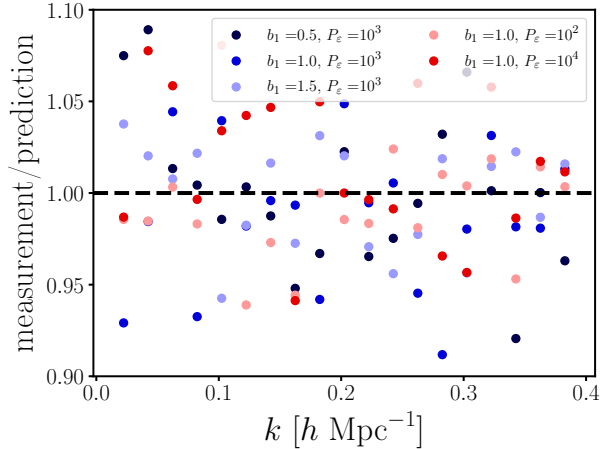}}
    \caption{Ratio of the variance of $10^3$ power spectra with fixed phases to the prediction obtained in Eq.~(\ref{eq: prediction variance}) for a variety of bias and noise parameters.}
    \label{Fixed Phases}
\end{figure}

In this section we derive the power spectrum covariance in the case where the initial phases are fixed. 
Inside a thin shell bin of magnitude $k$, which we keep fixed throughout, the prediction for the halo power spectrum can be written as 
\begin{equation}
    P_h(k) = \frac{1}{m_k}\sum_{\textbf{q}}|\delta_{\rm det}(\textbf{q}) + \varepsilon(\textbf{q})|^2,
\end{equation}
with the sum running over all the modes $\textbf{q}$ inside the bin of magnitude $k$. $\delta_{\rm det}(\textbf{q}) \equiv \delta_{\rm det}(\textbf{q}|\delta_{\rm in},\{b_{O}\})$ is the deterministic halo density field (for fixed phases $\delta_{\rm in}$) which can be found using the forward model, $\varepsilon(\textbf{q})$ is the noise field and $m_{k}$ is the number of modes inside that bin. We are interested in the variance of $P_h(k)$, i.e.,
\begin{align}
    \text{Var}_{\rm fix}[P
    _h(k)] &= \langle P_h^2(k) \rangle - \langle P_h(k) \rangle^2 \\
    &= \frac{1}{m_k^2} \sum_{\textbf{q}, \textbf{q}'}^{||\textbf{q}, \textbf{q}'|-k| < \Delta k/2} \left(\langle |\delta_h(\textbf{q})|^2|\delta_h(\textbf{q}')|^2 \rangle - \langle |\delta_h(\textbf{q})|^2\rangle\langle|\delta_h(\textbf{q}')| \rangle^2\right),
    \label{var}
\end{align}
in the case where $\delta_{\rm det}(\textbf{k})$ is fixed. 
We start by focusing on the right-hand side of Eq. \eqref{var}. The expected value $\langle |\delta_h(\textbf{q})|^2 \rangle$ for a single mode $\textbf{q}$ can be written as 
\begin{equation}
    \langle |\delta_h(\textbf{q})|^2 \rangle = \int \mathcal{D}\varepsilon \, \mathcal{P}(\varepsilon | P_\varepsilon) \, |\delta_{\rm det}(\textbf{q}) + \varepsilon(\textbf{q})|^2\,,
    \label{eq: delta square}
\end{equation}
where $\mathcal{P}(\varepsilon| P_\varepsilon)$ is a multivariate Gaussian given by
\begin{equation}
    \mathcal{P}(\varepsilon | P_\varepsilon) = \frac{1}{\sqrt{(2\pi P_\varepsilon)^{m_k}}} \, \text{exp} \left[ -\frac{1}{2} \sum^{m_k}_{\textbf{p}} \frac{|\varepsilon(\textbf{p})|^2}{P_\varepsilon}\right]
\end{equation}
and $P_\varepsilon\propto\sigma_\varepsilon^2$ is the noise power spectrum. Notice that in Eq. \eqref{eq: delta square} we integrate only over $\varepsilon$, since the value of $\delta_{\rm det}$ is fixed. 
Inserting
\begin{equation}
    |\delta_{\rm det}(\textbf{q}) + \varepsilon(\textbf{q})|^2 =  |\delta_{\rm det}(\textbf{q})|^2 + 2  \text{Re}[\delta_{\rm det}(\textbf{q}) \varepsilon^{*}(\textbf{q})] + |\varepsilon(\textbf{q})|^2
    \label{eq: sum_squared}
\end{equation}
in the integral of Eq. \eqref{eq: delta square}, only the first two terms will survive. The last term integrates to zero since $\mathcal{P}(\varepsilon | P_\varepsilon)$ is a symmetric function. Therefore, Eq. \eqref{eq: delta square} becomes
\begin{align}
    \langle |\delta_h(\textbf{q})|^2 \rangle &= \frac{1}{\sqrt{(2\pi P_\varepsilon)^{m_k}}}  \int \mathcal{D}\varepsilon(\textbf{p}) \,  \text{exp} \left[ -\frac{1}{2} \sum^{m_k}_{\textbf{p}\neq \textbf{q}} \frac{|\varepsilon(\textbf{p})|^2}{P_\varepsilon}\right] \\
    &\times \int \mathcal{D}\varepsilon(\textbf{q}) \,  \text{exp} \left[ -\frac{1}{2} \frac{|\varepsilon(\textbf{q})|^2}{P_\varepsilon}\right] \left(\,|\delta_{\rm det}(\textbf{q})|^2  +  |\varepsilon(\textbf{q})|^2 \right)\,.
    \label{firstwick}
\end{align}
This allows us to perform the integration for a single mode $\textbf{q}$ and obtain
\begin{equation}\label{eq: v1}
    \langle |\delta_h(\textbf{q})|^2 \rangle
    =  \,|\delta_{\rm det}(\textbf{q})|^2  + P_\varepsilon \,.
\end{equation}
This holds equivalently for $\textbf{q}'$, while the result for the whole bin can be found by summing over all the modes. 
Now let us focus on the first term in Eq. \eqref{var}, $\langle P_h^2(k) \rangle$,
where $\langle|\delta_h(\textbf{q})|^2|\delta_h(\textbf{q}')|^2\rangle=\langle|\delta_{\rm det}(\textbf{q}) + \varepsilon(\textbf{q})|^{2} |\delta_{\rm det}(\textbf{q}') + \varepsilon(\textbf{q}')|^{2}\rangle$ can be expanded as 
\begin{align}
    \langle|\delta_h(\textbf{q})|^2|\delta_h(\textbf{q}')|^2\rangle =\:&  \langle( |\delta_{\rm det}(\textbf{q})|^2 + 2  \text{Re}[\delta_{\rm det}(\textbf{q}) \varepsilon^{*}(\textbf{q})] + |\varepsilon(\textbf{q})|^2)\vs &\times (
     |\delta_{\rm det}(\textbf{q}')|^2 + 2  \text{Re}[\delta_{\rm det}(\textbf{q}') \varepsilon^{*}(\textbf{q}')] + |\varepsilon(\textbf{q}')|^2)\rangle \vs
    =\:& \langle  |\delta_{\rm det}(\textbf{q})|^2 |\delta_{\rm det}(\textbf{q}')|^2 +  |\delta_{\rm det}(\textbf{q})|^2 |\varepsilon(\textbf{q}')|^2 +  |\varepsilon(\textbf{q})|^2 |\delta_{\rm det}(\textbf{q}')|^2  
    \label{eq: first line} \\
    &+ |\varepsilon(\textbf{q})|^2 |\varepsilon(\textbf{q}')|^2+ 4\text{Re}[\delta_{\rm det}(\textbf{q}) \varepsilon^{*}(\textbf{q})]\text{Re}[\delta_{\rm det}(\textbf{q}') \varepsilon^{*}(\textbf{q}')] \rangle\,. 
    \label{eq: last line}
\end{align}
From the previous calculation of $\langle |\delta_h(\textbf{q})|^2 \rangle$, we already know how to calculate the expected values encountered in Eq. \eqref{eq: first line}. What is left for us to understand are the ones shown in the last line, Eq. \eqref{eq: last line}. Regarding the first term, since $\varepsilon$ is a random Gaussian field, by Wick's theorem we obtain
\begin{align}
    \langle \varepsilon(\textbf{q})\varepsilon^*(\textbf{q})\varepsilon(\textbf{q}')\varepsilon^*(\textbf{q}') \rangle =\:& \langle \varepsilon(\textbf{q})\varepsilon^*(\textbf{q}) \rangle\langle\varepsilon(\textbf{q}')\varepsilon^*(\textbf{q}') \rangle\vs 
    &+ \langle \varepsilon(\textbf{q}) \varepsilon(\textbf{q}') \rangle\langle\varepsilon^*(\textbf{q})\varepsilon^*(\textbf{q}') \rangle + \langle \varepsilon(\textbf{q}) \varepsilon^*(\textbf{q}') \rangle\langle\varepsilon^*(\textbf{q})\varepsilon(\textbf{q}') \rangle \vs
    =\:& P_\varepsilon^2 (1 + \delta_{\textbf{q},-\textbf{q}'} + \delta_{\textbf{q},\textbf{q}'})\,.
\end{align}
Regarding the last term from Eq. \eqref{eq: last line}, we can expand it as 

\begin{align}
    \langle 4\text{Re}[\delta_{\rm det}(\textbf{q}) \varepsilon^{*}(\textbf{q})]\text{Re}[\delta_{\rm det}(\textbf{q}') \varepsilon^{*}(\textbf{q}')] \rangle&= \langle[\delta_{\rm det}(\textbf{q})\varepsilon^*(\textbf{q}) + \delta^*_{\rm det}(\textbf{q}) \varepsilon(\textbf{q})][\delta_{\rm det}(\textbf{q}')\varepsilon^*(\textbf{q}') + \delta^*_{\rm det}(\textbf{q}') \varepsilon(\textbf{q}')]\rangle\vs
    &=
    \big\langle \delta_{\rm det}(\textbf{q})\varepsilon^*(\textbf{q})\delta_{\rm det}(\textbf{q}')\varepsilon^*(\textbf{q}') + \delta_{\rm det}(\textbf{q})\varepsilon^*(\textbf{q})\delta^*_{\rm det}(\textbf{q}')\varepsilon(\textbf{q}') 
    \label{eq: first real}\\
    &+ \delta^*_{\rm det}(\textbf{q})\varepsilon(\textbf{q})\delta_{\rm det}(\textbf{q}')\varepsilon^*(\textbf{q}') + \delta^*_{\rm det}(\textbf{q})\varepsilon(\textbf{q})\delta^*_{\rm det}(\textbf{q}')\varepsilon(\textbf{q}')  \big\rangle\,. \label{eq: last real}
\end{align}
Let us inspect how to calculate the expectation value of the first contribution of Eq. \eqref{eq: first real},
\begin{align}
    \langle \delta_{\rm det}(\textbf{q})\varepsilon^*(\textbf{q})\delta_{\rm det}(\textbf{q}')\varepsilon^*(\textbf{q}') \rangle &=\frac{1}{\sqrt{(2\pi P_\varepsilon)^{m_k}}}  \int \mathcal{D}\varepsilon(\textbf{p}) \,  \text{exp} \left[ -\frac{1}{2} \sum^{m_k}_{\textbf{p}\neq \textbf{q}} \frac{|\varepsilon(\textbf{p})|^2}{P_\varepsilon}\right]\vs
    &\times \int \mathcal{D}\varepsilon(\textbf{q}) \,  \text{exp} \left[ -\frac{1}{2} \frac{|\varepsilon(\textbf{q})|^2}{P_\varepsilon}\right] \, \delta_{\rm det}(\textbf{q})\varepsilon^*(\textbf{q})\delta_{\rm det}(\textbf{q}')\varepsilon^*(\textbf{q}')\vs
    &=|\delta_{\rm det}(\textbf{q})|^2 P_\varepsilon\, \delta_{\textbf{q},-\textbf{q}'}\,.
\end{align}
The calculation for the other three contributions of Eqs. \eqref{eq: first real}--\eqref{eq: last real} follows similarly. Collecting terms, we have that
\begin{align}
    \langle|\delta_h(\textbf{q})|^2|\delta_h(\textbf{q}')|^2\rangle 
    &=  |\delta_{\rm det}(\textbf{q})|^2 |\delta_{\rm det}(\textbf{q}')|^2 +  |\delta_{\rm det}(\textbf{q})|^2 P_\varepsilon +  |\delta_{\rm det}(\textbf{q}')|^2 P_\varepsilon \vs
    &+ P_\varepsilon^2 (1 + \delta_{\textbf{q},-\textbf{q}'} + \delta_{\textbf{q},\textbf{q}'}) + 2  |\delta_{\rm det}(\textbf{q})|^2 P_\varepsilon (\delta_{\textbf{q},-\textbf{q}'} + \delta_{\textbf{q},\textbf{q}'})
    \label{eq: v2}
\end{align}
By inserting Eq. \eqref{eq: v2} and the values of Eq. \eqref{eq: v1} for $\textbf{q}$ and $\textbf{q}'$ into the expression for the variance of the power spectrum given by Eq. \eqref{var}, we find that
\begin{align}
    \text{Var}_{\rm fix}[P
    _h(k)] &= \frac{1}{m_k^2} \sum_{\textbf{q}, \textbf{q}'}^{||\textbf{q}, \textbf{q}'|-k| < \Delta k/2} \left[\langle |\delta_h(\textbf{q})|^2|\delta_h(\textbf{q}')|^2 \rangle - \langle |\delta_h(\textbf{q})|^2\rangle\langle|\delta_h(\textbf{q}')| \rangle^2\right]\vs
    = \frac{1}{m_k^2} \sum_{\textbf{q}, \textbf{q}'}^{||\textbf{q}, \textbf{q}'|-k| < \Delta k/2}&
    \big[ |\delta_{\rm det}(\textbf{q})|^2 |\delta_{\rm det}(\textbf{q}')|^2 +  P_\varepsilon (|\delta_{\rm det}(\textbf{q})|^2 + |\delta_{\rm det}(\textbf{q}')|^2) + P_\varepsilon^2 \vs
    &+ (P_\varepsilon^2  + 2 |\delta_{\rm det}(\textbf{q})|^2 P_\varepsilon) \, (\delta_{\textbf{q},-\textbf{q}'} + \delta_{\textbf{q},\textbf{q}'})- ( |\delta_{\rm det}(\textbf{q})|^2 + P_\varepsilon)( |\delta_{\rm det}(\textbf{q}')|^2 + P_\varepsilon)\big]\,.
\end{align}
Summing over $\textbf{q}'$, we finally obtain the power-spectrum variance when the phases of $\delta$ are fixed,
\begin{equation}
    \text{Var}_{\rm fix}[P
    _h(k)] = \frac{2 P_\varepsilon}{m_k^2} \sum_{\textbf{q}}^{||\textbf{q}|-k| < \Delta k/2} (2 |\delta_{\rm det}(\textbf{q})|^2 + P_{\varepsilon})\,.
    \label{eq: prediction variance}
\end{equation}
Using a mock generator for $\delta_{\rm det}=b_1\delta$ in which the phases of $\delta$ are fixed, we verify in Fig. \ref{Fixed Phases} that this prediction accurately matches the variance of $10^3$ power spectra measured in a $512h^{-1}$Mpc box with $128^3$ cells. 

\pagebreak

\bibliographystyle{unsrt.bst}
\bibliography{References}

\end{document}